\documentclass[sn-mathphys-num,pdflatex]{sn-jnl}%

\usepackage{graphicx}%
\usepackage{amsmath,amssymb,amsfonts}%
\usepackage{amsthm}%
\usepackage{mathrsfs}%
\usepackage[title]{appendix}%
\usepackage{xcolor}%
\usepackage{textcomp}%
\usepackage{manyfoot}%
\usepackage{algorithm}%
\usepackage{algorithmicx}%
\usepackage{algpseudocode}%
\usepackage{listings}%
\usepackage{float}
\usepackage{subcaption}

\usepackage{hyperref}

\usepackage{tikz}
\usepackage{pgfplots}
\pgfplotsset{compat=1.9}
\usetikzlibrary{colorbrewer}
\usepgfplotslibrary{colormaps}
\usetikzlibrary{pgfplots.colormaps}
\usepackage{pgf-pie}

\usepackage{rotating}

\usepackage{float}
\usepackage{booktabs}
\usepackage{tabularx}
\usepackage{comment}
\usepackage{enumitem}
\usepackage{pdflscape}
\usepackage{wrapfig}
\usepackage{longtable}
\usepackage{tablefootnote}
\usepackage{tabularray}
\usepackage{array}\newcolumntype{C}[1]{>{\centering\arraybackslash}p{#1}}
\usepackage{makecell}
\usepackage{multirow}

\newcommand\TB{\textbullet}

\usepackage{ragged2e}

\usepackage{caption}
\usepackage{xcolor}
\usepackage{chronology}
\usepackage{marginnote}

\pgfplotsset{colormap={myBlues}{
    rgb255=(247,251,255)
    rgb255=(222,235,247)
    rgb255=(198,219,239)
    rgb255=(158,202,225)
    rgb255=(107,174,214)
    rgb255=(66,146,198)
    rgb255=(33,113,181)
    rgb255=(8,81,156)
    rgb255=(8,48,107)
}}

\pgfplotsset{colormap={myReds}{
    rgb255=(255,245,240)
    rgb255=(254,224,210)
    rgb255=(252,187,161)
    rgb255=(252,146,114)
    rgb255=(251,106,74)
    rgb255=(239,59,44)
    rgb255=(203,24,29)
    rgb255=(165,15,21)
    rgb255=(103,0,13)
}}

\begin{document}

\title[Reviewing Uses of Regulatory Compliance Monitoring]{Reviewing Uses of Regulatory Compliance Monitoring}

\author*[1,2]{\fnm{Finn} \sur{Klessascheck}}\email{finn.klessascheck@tum.de}

\author[1,2]{\fnm{Luise} \sur{Pufahl}}\email{luise.pufahl@tum.de}

\affil*[1]{\orgdiv{School of CIT}, \orgname{Technical University of Munich}, \orgaddress{\street{Bildungscampus 2}, \city{Heilbronn}, \postcode{74076}, \country{Germany}}}

\affil[2]{\orgname{Weizenbaum Institute}, \orgaddress{\street{Hardenbergstraße 32}, \city{Berlin}, \postcode{10623}, \country{Germany}}}

\abstract{
Organizations need to manage numerous business processes for delivering their services and products to customers. One important consideration thereby lies in the adherence to regulations such as laws, guidelines, or industry standards. In order to monitor adherence of their business processes to regulations -- in other words, their regulatory compliance -- organizations make use of various techniques that draw on process execution data of IT systems that support these processes. Previous research has investigated \emph{conformance checking}, an operation of process mining, for the domains in which it is applied, its operationalization of regulations, the techniques being used, and the presentation of results produced. However, other techniques for regulatory compliance monitoring, which we summarize as compliance checking techniques, have not yet been investigated regarding these aspects in a structural manner.
To this end, this work presents a systematic literature review on uses of regulatory compliance monitoring of business processes, thereby offering insights into the various techniques being used, their application and the results they generate. %
We highlight commonalities and differences between the approaches and find that various steps are performed manually; we also provide further impulses for research on compliance monitoring and its use in practice.

}


\maketitle
\vspace{-3em}
\textbf{Acknowledgements} We thank Ilona Bogatinovska and Tom Knoche for their contributions towards this article.

\vspace{-0.1em}
\noindent
\textbf{Funding Information} Funded by the Deutsche Forschungsgemeinschaft (DFG, German Research Foundation), grant no. 465904964.

\section{Introduction}
\label{sec:intro}

In order to deliver their services and products to customers, organizations must manage a broad range of \emph{business processes}~\cite{dumas2013fundamentals}.
These business processes can be understood as structured and repeatable sets of activities designed to achieve specific business objectives~\cite{Weske19}. In practice, these business processes are usually subject to \emph{regulations}, such as laws, industry standards, or guidelines~\cite{governatoriJourneyBusinessProcess2009}. These regulations originate from the processes environment and govern their execution by imposing constraints upon them.
For businesses to achieve organizational success, it is important to comply with relevant regulations~\cite{lopezBusinessProcessCompliance2020}, as not doing so could bring about unintended consequences, such as fines or loss of reputation~\cite{klessascheckReviewingConformance2024}.

When aiming to ensure that all relevant regulations are followed, organizations usually face two options: They can either assess the compliance of their business processes when they are \emph{designed}~\cite{hashmiAreWeDone2018} (so that an implementation is automatically compliant), or \emph{monitor} their compliance during and after \emph{execution}~\cite{groefsemaUseConformanceCompliance2022}. Notably, compliance monitoring brings the possibility of designing process execution in a more flexible manner, while still ensuring compliance, in comparison to a process that is compliant by design but must be strictly adhered to during execution to not violate compliance.
One particular technique, \emph{conformance checking}~\cite{carmona2018conformance}, has been developed in the fields of \emph{business process management} (BPM) and \emph{process mining}, which has been identified as a useful technique for regulatory compliance monitoring~\cite{klessascheckReviewingConformance2024}: With conformance checking, the conformance of real-world process execution to desired process behaviour can be assessed in a data-driven manner.

Existing research has already investigated conformance checking uses for regulatory compliance monitoring~\cite{klessascheckReviewingConformance2024}; however, other techniques for checking compliance, such as those based on \emph{complex event processing} also exist and can be applied for monitoring business process compliance~\cite{lyComplianceMonitoringBusiness2015}.
While a wide range of research on \emph{applications} of compliance monitoring has been conducted (e.g., case studies that use conformance checking~\cite{lemosUsingProcessMining2011}, contributions that investigate the potential for regulatory compliance monitoring with conformance checking more abstractly~\cite{klessascheckUnlocking2024}), 
research currently does not provide a general overview of \emph{usages} of various techniques available for monitoring regulatory compliance of business processes. To provide such an overview, thereby increasing awareness of existing approaches and identifying the potential for further contributions, we conduct a systematic literature review. In doing so, this work aims to answer the following research questions:

\medskip
\begin{enumerate}[label=\textbf{RQ\arabic*:},wide=0pt, leftmargin=*]
	\item What are existing peer-reviewed studies that practically apply compliance monitoring techniques to assess the regulatory compliance of business processes?
	\item What characterizes the actual compliance monitoring approaches in the contributions?
	\item What are, based on the characterization of existing studies, potential areas of further investigation for the application of compliance monitoring techniques?
\end{enumerate}

\medskip
This paper is based on a previous study, where we reviewed conformance checking uses for regulatory compliance monitoring purposes~\cite{klessascheckReviewingConformance2024}. Extending our focus, we now additionally investigate usages of further techniques that exist in the research space of regulatory compliance and BPM in our review, and take further aspects such as the \emph{origin} and \emph{availability} of process execution data used in the investigated studies into account. We also describe our research method in more detail and, in light of our broadened scope, extend the results and discussion provided herein.

The remainder of this article is organized as follows: In Sec.~\ref{sec:background}, we provide background on compliance monitoring in general and conformance checking in particular, as well as related work. In Sec.~\ref{sec:methods} we present our research method for identifying and synthesizing relevant studies. In Sec.~\ref{sec:results-analysis}, we provide our analysis of the relevant studies, while we discuss our findings and the overall contributions in Sec.~\ref{sec:discussion}. We conclude in Sec.~\ref{sec:concl}, where we also present potential for future work.

\section{Background and Related Work}
\label{sec:background}

In the following, we provide background on compliance monitoring, and discuss related work that has investigated these techniques and their applications. 

\subsection{Compliance Monitoring}
\label{sec:background-compiance-monitoring}
As discussed above, it is necessary for businesses to assess whether their behaviour complies with the regulations to which they are subject. 
This act of ensuring that business processes do not violate the regulations relevant for them is called \emph{regulatory compliance checking}~\cite{groefsemaUseConformanceCompliance2022}. One existing type of approach, being regulatory \emph{compliance monitoring} (also known in the literature as \emph{run-time regulatory compliance checking}~\cite{groefsemaUseConformanceCompliance2022}), aims at monitoring process executions to assess their compliance with regulations~\cite{vanbeestCrossInstanceRegulatoryCompliance2023}.
This type of approach contrasts with traditional \emph{design-time regulatory compliance checking}~\cite{groefsemaUseConformanceCompliance2022}, in which the process model itself is analysed for violations of and compliance with regulations. However, processes can in practice be executed in ways that differ from the process model, thus giving relevancy to compliance monitoring during run-time~\cite{hashmiAreWeDone2018}. Note that while the compliance of business processes with business rules, that is, rules which embody organizational practice or policy~\cite[Ch.4]{dumas2013fundamentals}, can be assessed with compliance monitoring, we here focus specifically on \emph{regulatory} compliance. This means assessing whether business processes comply with what law or rules ask or order them to do~\cite{groefsemaUseConformanceCompliance2022}. While other (system) requirements exist that need to be checked against business process models and executions (such as user requirements and design properties~\cite{groefsemaUseConformanceCompliance2022}), the importance of regulations that impose constraints on the behaviour of business processes, and the need of proving that these constraints are adhered to, motivates the focus of this study. Here, we concentrate on regulatory compliance of business processes specifically, instead of verifying the behaviour of a system and its implementation against a business process model more generally.

\begin{figure}[ht!]
    \centering
    \includegraphics[width=0.8\textwidth]{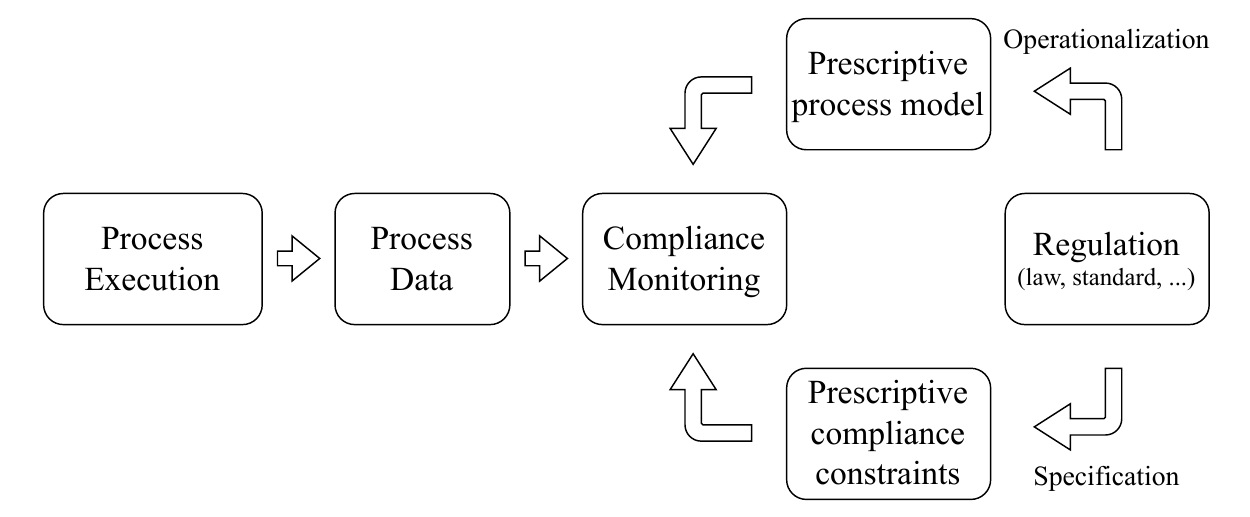}
    \caption{Conceptual overview of compliance monitoring with conformance checking or compliance monitoring, adapted from~\cite{klessascheckUnlocking2024,lyComplianceMonitoringBusiness2015}}
    \label{fig:overview-compliance-monitoring}
\end{figure}

In research, various techniques for regulatory compliance monitoring have been identified as useful, among them conformance checking~\cite{klessascheckReviewingConformance2024}. In the following, we provide, as background information, a brief overview of the most salient techniques. Figure~\ref{fig:overview-compliance-monitoring} provides a conceptual overview of compliance monitoring. 

In general, regulations are usually available in the form of textual documents~\cite{klessascheckReviewingConformance2024}. Hence, for monitoring regulatory process compliance, a specification, or \emph{operationalization}, becomes necessary to relate the regulation to the business process under investigation. Depending on the technique and approach, which we will further elaborate upon in the following, different formal representations are created: Regulatory constraints relevant for a business process are operationalized, that is, used to create prescriptive process models or specific formal process constraints. These are subsequently used in various compliance monitoring techniques. Regardless of techniques and the type of approach (i.e., whether design-time or run-time regulatory compliance is checked with specific techniques), these operationalized regulations are intermediate, explicitly modelled, representations that encapsulate regulatory constraints, and are therefore an important component when assessing regulatory compliance of business processes~\cite{groefsemaUseConformanceCompliance2022}.

\subsubsection{Conformance Checking}

Central to the execution of business processes are information systems, which capture information about the processes and their execution in \emph{event logs}. These event logs, containing ordered records of individual process instances and the events that record what activity has been executed when~\cite{van2012process}, are the basis of conformance checking: The process event data of event logs can be linked to a process model, allowing insights into the relation between observed and modelled behaviour~\cite{carmonaConformanceCheckingFoundations2022}.
Commonly considered with conformance checking is the question of whether process behaviour captured in an event log conforms to prescribed (normative) behaviour. This normative behaviour is given in the form of a \emph{prescriptive} process model~\cite{carmona2018conformance,groefsemaUseConformanceCompliance2022}. Process models are (mainly graphical) representations of business processes, and describe the process activities and their ordering~\cite{Weske19}.
Consequently, techniques from the family of \emph{conformance checking} can be used to assess the relation of recorded and prescribed process behaviour. Conformance checking techniques usually express the result of such an analysis in the form of fitness values, i.e., the degree of conformance between event log and process model, and provide lists of deviations, i.e., recorded process executions that deviate from the process models. Different techniques for conformance checking exist, mainly differing in the type of prescriptive model they use~\cite{dunzerConformanceCheckingStateoftheart2019}. \emph{Imperative} techniques use imperative process models, such as BPMN~\cite{bpmnspec} or Petri nets, which describe only allowed behaviour. \emph{Declarative} techniques apply declarative process models, such as Declare~\cite{pesicDECLAREFullSupport2007}, which only express constraints on process behaviour, instead of explicitly specifying it.

Further, \emph{hybrid} techniques use a mixture of imperative and declarative elements to prescribe behaviour of business processes~\cite{van2021conformance}. In general, prescriptive models formalize constraints towards one or more \emph{process perspectives}, which are characteristics of processes relevant for conformance checking. These perspectives include the ordering of activities, temporal aspects, data and documents, and resources that execute the process~\cite{russellWorkflowPatternsDefinitive2015}.

In practice, conformance checking has been identified as a suitable and useful approach for regulatory compliance monitoring~\cite{carmona2018conformance}. Here, regulations relevant for a business process are interpreted and operationalized into prescriptive process models that express compliance constraints. These prescriptive models are then used to monitor the regulatory compliance of recorded process event data in the form of event logs, with the help of conformance checking techniques. This is done either based on event logs (offline and ex-post, similar to auditing), or based on event streams (online, during run-time).
According to Groefsema et al.~\cite{groefsemaUseConformanceCompliance2022}, the main issue with using conformance checking for regulatory compliance monitoring lies in the fact that the prescriptive model being used must \emph{also} be proven to be regulatory compliant with regulatory compliance checking. However, given that that conformance checking \emph{is} indeed applied for regulatory compliance monitoring, a systematic investigation is warranted.

\subsubsection{Compliance Checking}

While conformance checking techniques can be used for regulatory compliance monitoring, other techniques that do not belong to the family of conformance checking exist. Here, we summarize them under the umbrella term of \emph{compliance checking}, in line with~\cite{groefsemaUseConformanceCompliance2022,lyComplianceMonitoringBusiness2015}.
\emph{Complex event processing} (CEP)~\cite{luckhamPowerEvents2008} in particular, focussed on processing and aggregating streams of process event data, has been positioned as an enabling technique for regulatory compliance monitoring~\cite{lyComplianceMonitoringBusiness2015,knupleschAFramework2017,hashmiAreWeDone2018}: In CEP, process event data generated during process execution is collected and aggregated into higher-level events, to which monitoring rules and constraints are applied so that non-compliant behaviour can be detected~\cite{knupleschVisuallyMonitoring2015,hashmiAreWeDone2018}. In general, CEP can be used for monitoring various types of system requirements, such as user requirements or design properties regarding e.g. information flow of web services~\cite{fischerSecurityPolicyEnforcement2007}; the focus of using CEP for compliance monitoring of business processes, and thus of this study, rather lies on the application of techniques for monitoring process events that make up business process executions~\cite{lyComplianceMonitoringBusiness2015}. These business processes in turn are subject to compliance constraints; in the case of regulatory compliance monitoring, these constraints are representations of regulatory constraints. We therefore do not consider applications of approaches that focus on granularities beyond those of business process instances, such as system requirements or service requirements, e.g. SLAs, or compliance constraints that do not stem from regulations.

Regardless of the technique, in compliance checking approaches, it is also necessary to interpret compliance requirements and specify compliance constraints~\cite{lyComplianceMonitoringBusiness2015}; these constraints, similar to a prescriptive model of conformance checking techniques, relate to process perspectives and could be imperative, declarative, or hybrid representations of prescribed behaviour~\cite{lyComplianceMonitoringBusiness2015}. For example, declarative representations include the \emph{extended Compliance Rule Graph Language} (eCRG) which consists of control flow, data, resources and temporal elements, with which regulatory constraints can be expressed in prescriptive models~\cite{hashmiAreWeDone2018,knupleschVisuallyMonitoring2015}; further examples include linear-temporal logic (LTL) and Declare~\cite{hashmiAreWeDone2018}. Imperative techniques for compliance checking might use, similar to conformance checking, traditional imperative process models from which then monitoring queries for use in CEP environments can be derived~\cite{weidlichEventBasedMonitoring2011}. 
For a more comprehensive \emph{technical} overview of compliance checking, we refer to the work of Hashmi et al.~\cite{hashmiAreWeDone2018}.

In the literature, compliance checking has been characterized as being more suitable for monitoring regulatory compliance than conformance checking. As argued by~\cite{hashmiAreWeDone2018,hashmiNormativeRequirementsRegulatory2016}, this is because the imperative models typically used for conformance checking would primarily express \textit{business rules} and not \textit{regulatory constraints}~\cite{hashmiAreWeDone2018,groefsemaUseConformanceCompliance2022,hashmiNormativeRequirementsRegulatory2016}. Business rules are here understood as ``declarations which constrain, derive and give conditions for existence, representing the knowledge of the business, [… and] define the conditions under which a process is carried out or the new conditions that will exist after a process has been completed''~\cite{araujoAMethod2010}, whereas regulatory constraints describe conditions and obligations stemming directly from the legal domain~\cite{hashmiNormativeRequirementsRegulatory2016,hashmiAreWeDone2018}. Still, existing research also underlines similarities and synergies between conformance and compliance checking approaches~\cite{lyComplianceMonitoringBusiness2015}. Further, the difference between conformance and compliance checking approaches has been blurred due to terminological inconsistencies: in fact, checking whether observed behaviour follows a specification, is in the strictest sense \emph{conformance} checking, even if the goal is to prove regulatory \emph{compliance}~\cite{groefsemaUseConformanceCompliance2022}. Notably, we see conformance checking applied in practice for regulatory compliance monitoring purposes, also using declarative techniques and prescriptive models~\cite{klessascheckReviewingConformance2024}); we are further aware of studies that position themselves as checking compliance directly. Therefore, it appears prudent to further investigate similarities and differences in practical uses of compliance monitoring as reported in studies that position themselves as applying compliance and conformance checking.

\subsection{Related Work}

Various contributions have already aimed at systematically assessing the capabilities of existing compliance monitoring approaches, in particular compliance checking and conformance checking techniques.
Ly et al.~\cite{lyComplianceMonitoringBusiness2015} present a framework for systematic comparison of compliance monitoring approaches along modelling requirements, execution requirements and user requirements covered by existing tooling. Modelling requirements include the ability to model time, data, and resource constraints, whereas user requirements include the ability to reactively or proactively detect violations, the ability to explain the root cause of deviations, and the ability to quantify the degree of compliance. Execution requirements relate to the ability of dealing with information available during execution of a business process on an event level, such as dealing with activities that are non-atomic, checking for constraints of an activity' life cycle, or supporting multiple instances of the same constraint in a single process instance. Existing compliance monitoring approaches are classified according to whether they fulfil these requirements by applying all of them to one particular example. 

Building on this framework, Rinderle-Ma et al.~\cite{rinderle-maPredictiveComplianceMonitoring2023} classify \emph{predictive} compliance monitoring approaches along compliance monitoring requirements and capabilities. For doing so, they extend the requirements formulated by Ly et al.~\cite{lyComplianceMonitoringBusiness2015} with e.g. additional user requirements or data requirements specific to \emph{predictive} process compliance monitoring. Similar to Ly et al.~\cite{lyComplianceMonitoringBusiness2015}, they assess in how far existing literature covers the functionality requirements they elicited, and derive research directions from this assessment. %
Requirements towards compliance monitoring approaches are elicited from real-world case studies and literature in both studies; however, concrete applications of these approaches are not characterized and investigated for their particular properties. While the requirements are useful for describing the capabilities of compliance monitoring approaches, they are insufficient to characterize \emph{how} these approaches are \emph{applied} in a way that goes beyond their ability to meet e.g. modelling, execution, and user requirements. Instead, we believe more emphasis should be placed on \emph{how} prescriptive models come to be, to what end uses of approaches are reported, and in what ways the results are represented (see also e.g. Rehse et al.~\cite{rehse2022process}), in particular for regulatory compliance monitoring.

Similarly, the literature on conformance checking has focussed on technical aspects exclusively, and less on conformance checking approaches. Dunzer et al.~\cite{dunzerConformanceCheckingStateoftheart2019} analyse conformance checking techniques for the modelling language and the type of algorithm being used, the kind of metric with which conformance is expressed, and the process perspective that is considered (e.g., the ordering of activities or temporal aspects). However, the work does not focus on compliance checking with regulations in particular, nor is the creation of the prescriptive models analysed.
Further, Oliart et al.~\cite{OLIART2022104076} exclusively investigate approaches that assess clinical guidelines. Concretely, the guidelines that are used by existing studies are analysed per disease, their complexity is assessed, as well as the prescriptive model in terms of, e.g., nodes in the process model or number of declarative constraints --- there is no further consideration of how these prescriptive models are created, nor what the results are or how the techniques are applied.

Thus, we see a need to investigate how applications of compliance monitoring use regulations in a practical sense beyond purely technical considerations, how they are applied, and what results they provide.

\section{Research Method}
\label{sec:methods}

To address the research questions above, we conduct a \emph{systematic literature review} (SLR) to identify scientific studies that apply a) conformance checking techniques and b) other techniques of the compliance space, for regulatory compliance monitoring of business processes with \emph{concrete} regulations. In this, our study constitutes a \emph{descriptive review}, as we aim to identify trends and patterns in the application of compliance monitoring techniques~\cite{pareSynthesizingInformationSystems2015}.
For this, we adhere to the eight-step method of Okoli~\cite{okoliGuideConductingStandalone2015}. These steps are as follows: (1) purpose of the literature review, (2) protocol and training, (3) searching for the literature, (4) practical screen, (5) quality appraisal, (6) data extraction, (7) synthesis of studies, and finally (8) writing the review. Additionally, we use a concept matrix as a framework for presenting the results~\cite{websterAnalyzingPrepareFuture2002}.

\begin{figure}[ht!]
    \centering
    \includegraphics[width=0.9\textwidth]{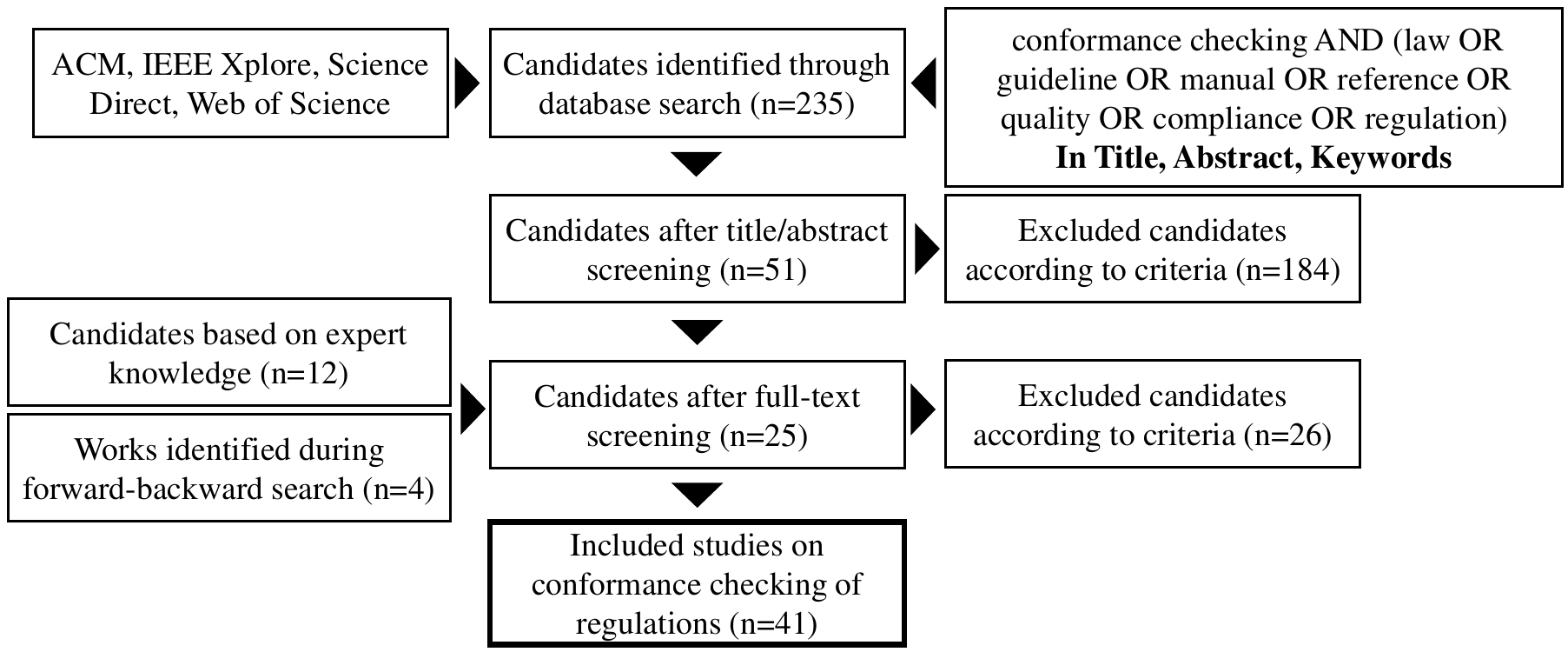}
    \caption{Search process of Search A for relevant literature, adapted from~\cite{klessascheckReviewingConformance2024}}
    \label{fig:search_process_a}
\end{figure}

\begin{figure}[ht!]
    \centering
    \includegraphics[width=0.9\textwidth]{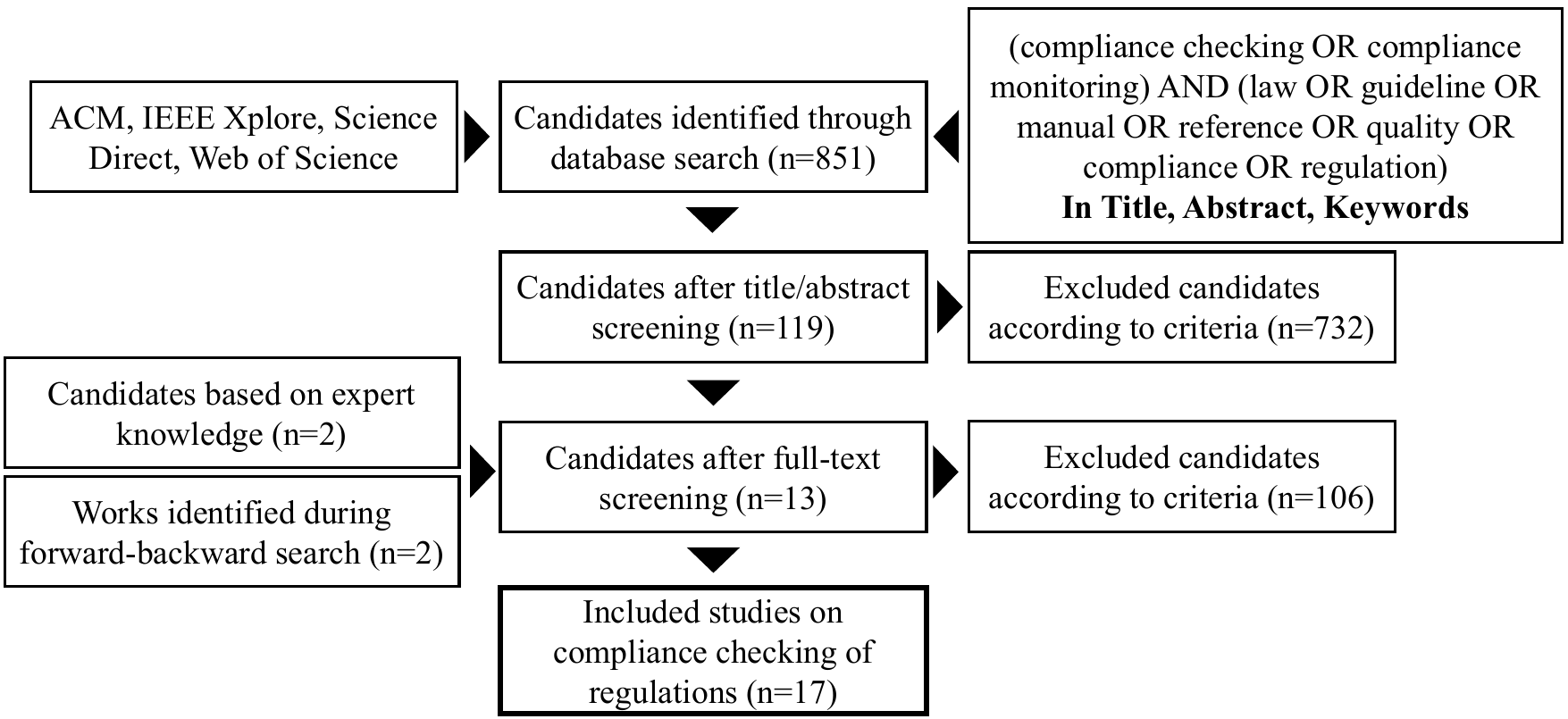}
    \caption{Search process of Search B for relevant literature, adapted from~\cite{klessascheckReviewingConformance2024}}
    \label{fig:search_process_b}
\end{figure}

For the \emph{purpose} of this SLR, as elaborated upon in Sec.~\ref{sec:intro}, we aim to assess the existing state of the art of approaches that apply regulatory compliance monitoring techniques, and to identify opportunities for further research.  %
For the \emph{protocol and training} step, we set up a shared document in which we documented all identified articles and reasons for their inclusion respective exclusion.
For the \emph{search} step, we conduct two separate searches using four scientific databases (ACM, IEEE Explore, Science Direct, Web of Science), up to and including May 2024. Conducting two separate searches allows us to better differentiate between studies that \emph{position themselves} as applying conformance checking techniques and as applying compliance checking techniques for regulatory compliance monitoring, thus allowing us to investigate commonalities and differences between these two perspectives on compliance monitoring. Figures~\ref{fig:search_process_a} and~\ref{fig:search_process_b} illustrate our search process.

The first search (subsequently referred to as \emph{Search A}, illustrated in Fig.~\ref{fig:search_process_a}), aims to identify contributions that apply conformance checking as a technique for regulatory compliance monitoring exclusively. The second search (referred to as \emph{Search B}, shown in Fig.~\ref{fig:search_process_b}), aims to identify contributions that apply any other technique (which we subsume as \emph{compliance checking} following~\cite{lyComplianceMonitoringBusiness2015,vanbeestCrossInstanceRegulatoryCompliance2023}) for regulatory compliance monitoring. Doing so allows us to differentiate between approaches that consider compliance monitoring from the conformance checking perspective, and those that come from the compliance checking perspective unrelated to conformance checking techniques.

In Search A, we explicitly focus on studies that position themselves as applying conformance checking techniques as an approach for regulatory compliance monitoring. Moreover, we include terms commonly associated with regulations. The complete search term \emph{(“conformance checking” AND (“law” OR “guideline”, “manual” OR “reference” OR “quality” OR “compliance” OR “regulation”))} is applied to the title, abstract, and keyword search, where applicable. In Web of Science, we filtered for the categories of computer science and business. %

In Search B, we explicitly search for studies that position themselves as applying techniques besides conformance checking (meaning, other compliance checking techniques) for regulatory compliance monitoring. In line with Search A, we include terms commonly associated with regulations. The complete search term \emph{((“compliance checking” OR “compliance monitoring”) AND (“law” OR “guideline” OR “manual” OR “reference” OR “quality” OR “compliance” OR “regulation”))} is applied to the title, abstract, and keyword search, where applicable.\footnote{Note that for the search we use compliance monitoring and compliance checking interchangeably for any study that mentions applications of compliance monitoring or checking techniques, due to a certain flexibility of the terms' use we observed in the literature.} In Web of Science, we filtered for the categories of computer science and business. %

Aiming to investigate the scientific support for practical compliance monitoring with conformance checking and other techniques, we have set up several exclusion (EX) and inclusion (IN) criteria to screen for relevant studies. In particular, we exclude (EX1) all studies that assess this topic in an abstract manner, such as literature reviews, (EX2) studies that provide general techniques or metrics for regulatory compliance monitoring without application to recorded process executions, and (EX3) studies where compliance monitoring techniques are not applied to a \emph{concrete} business process and regulation. We explicitly include (IN1) studies that apply compliance monitoring techniques to business processes (meaning, recorded process executions) against specific guidelines, laws, regulations, manuals, etc., (IN2) which are peer-reviewed and written in English.\footnote{We did not, however, exclude peer-reviewed studies based on the venue or journal in which they appeared, in order to capture a wide range of contributions, which is consistent with descriptive reviews~\cite{pareSynthesizingInformationSystems2015}.} This constitutes the \emph{practical screen}.

An initial Search A -- illustrated in Fig.~\ref{fig:search_process_a} -- with the databases and search term described above produces, after deduplication, 235 candidates for analysis. Afterward, we conduct the \emph{quality appraisal} by deciding, first based on the title and abstract and then on the full text, whether the found studies meet our criteria. In total, 51 remain after the title and abstract screening and 25 after the full-text screening. We add twelve papers through the authors' expert knowledge and identify four additional studies via a “forward-backward-search” we conduct based on the relevant papers of the full-text screening to find potentially missing studies. Both steps are in line with existing recommendations for conducting literature reviews to ensure that the search covers sufficient literature~\cite{okoliGuideConductingStandalone2015,vombrockeStandingShouldersGiants2015}. This leads to 41 papers we consider relevant for our analysis from Search A.

For Search B, illustrated in Fig.~\ref{fig:search_process_b}, we identify 851 candidates for analysis after an initial search and deduplication, as described above. Afterwards, similarly to Search A, we conduct the \emph{quality appraisal}; after title and abstract screening, 119 articles remain, and 13 after full-test screening. Through a “forward-backward-search”, we identify two additional relevant studies, and two via expert knowledge. Thus, Search B has produced 17 papers we consider relevant for our analysis.

As part of papers added via expert knowledge, we filter out studies in Search A that clearly apply compliance checking techniques instead of conformance checking techniques and add them to Search B as additional expert knowledge; similarly, studies that clearly apply conformance checking techniques instead of compliance checking are screened out and added to Search A as additional expert knowledge. This was done for five studies in total (one study from Search A to B, and four from B to A). For clarity, Figs.~\ref{fig:search_process_a} and~\ref{fig:search_process_b} show these studies as part of the ``expert knowledge'' step, and they have been already included in the preceding paragraphs. 
After having identified 41/17 (Search A/Search B) relevant studies, we continue with \emph{data extraction} and \emph{synthesis} of the extracted information for each of the two searches.

In relation to our research questions, we define a set of categories and corresponding characteristics to capture the distinctions within the compliance monitoring application of each study. We first extract detailed information for each category. Subsequently, we analyse the extracted information for similarities in their characteristics. Finally, as the synthesis step, we identify potential levels of abstraction that accurately reflect the information while avoiding excessive specificity. This approach is consistent with Okoli's~\cite{okoliGuideConductingStandalone2015} methodology, where our synthesis step is qualitative.%

\begin{figure}[ht!]
    \centering
    \includegraphics[width=0.8\linewidth]{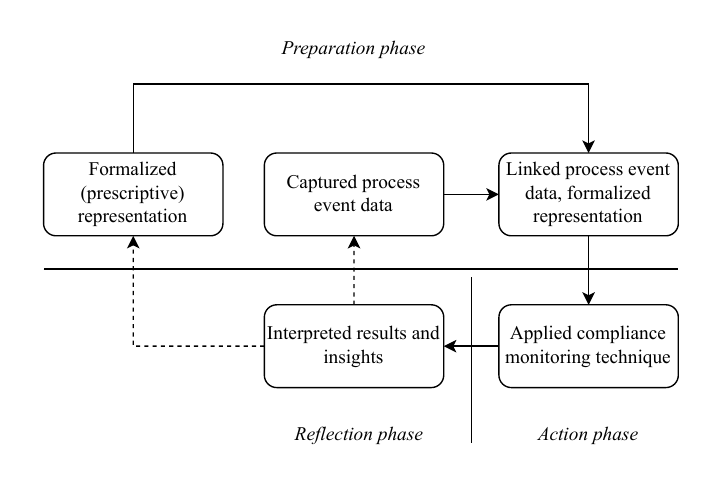}
    \caption{Lifecycle and phases of compliance monitoring, adapted from Carmona et al.'s lifecycle of conformance checking projects~\cite{carmona2018conformance}, used in this study to guide data extraction and synthesis}
    \label{fig:cc_cycle}
\end{figure}

For the results of Search A and Search B, we identify characteristics in relation to the lifecycle model of conformance checking projects of Carmona et al.~\cite{carmona2018conformance}, which divides the application of conformance checking into three phases: 1.) \emph{preparation}, where a prescriptive model is formalized, process event data is captured in an event log, and the two related; 2.) \emph{action}, where a compliance monitoring technique, such as conformance checking, is applied; and 3.) \emph{reflection}, where results are interpreted and insights derived. These steps are closely related to the generalized approach of compliance monitoring described by Ly et al.~\cite{lyComplianceMonitoringBusiness2015} (where compliance rules are formally specified or modelled, applied to process event data generated by process executions, and results are reported), although more structured with an explicit iteration; therefore we adopt the conformance checking lifecycle model as our theoretical lens. Figure~\ref{fig:cc_cycle} displays the adapted lifecycle of compliance monitoring. The results, in the form of a concept matrix following Webster and Watson~\cite{websterAnalyzingPrepareFuture2002}, can be found in Table~\ref{tab:coding}.

\begin{table}[ht!]
\caption{Concept matrix~\cite{websterAnalyzingPrepareFuture2002} displaying the synthesis results. \emph{PM} refers to \emph{prescriptive model}; the domains and results are abbreviated.\tablefootnote{The domains are \textbf{C}ustomer \textbf{S}ervices, \textbf{D}ata \textbf{P}rocessing, \textbf{Edu}cation, \textbf{Fin}ance, \textbf{H}ealth\textbf{c}are, \textbf{IT S}ervice \textbf{M}anagement, \textbf{H}uman \textbf{R}esources, \textbf{Manu}facturing, \textbf{Nuc}lear Power, \textbf{P}ublic \textbf{S}ervices, \textbf{Rent}, \textbf{Sci}ence, \textbf{S}oft\textbf{w}are, \textbf{Sup}ply Chain, \textbf{Trade}. Besides Qualify and Quantify, the results contain \textbf{B}reak\textbf{d}own and \textbf{Comp}are, \textbf{Loc}alize and \textbf{Show}, and \textbf{Expl}ain and \textbf{Diagn}ose.}
}
\centering
\renewcommand{\TB}{\textbullet}
 \setlength\tabcolsep{1pt}
     \tiny
    \begin{tabularx}{1\textwidth}{|l|l|X|X|X|X|X|X|X|X|X|X|X|X|X|X|X|X|X|X|X|X|X|X|X|X|X|X|X|X|}
    \hline
    & & & \multicolumn{10}{c|}{Preparation} & \multicolumn{12}{c|}{Action} & \multicolumn{5}{c|}{Reflection} \\\hline\hline
    \begin{sideways}Study\end{sideways} & \begin{sideways}Domain\end{sideways} & \begin{sideways}Search\end{sideways} & \multicolumn{3}{l|}{\begin{sideways}Goal\end{sideways}} & \multicolumn{2}{l|}{\begin{sideways}PM Creation\end{sideways}} & \begin{sideways}PM Validated\end{sideways} & \multicolumn{2}{l|}{\begin{sideways}Data Type\end{sideways}} & \multicolumn{2}{l|}{\begin{sideways}Data Aval.\end{sideways}} & \begin{sideways}Exp. Knowl.\end{sideways} & \multicolumn{3}{l|}{\begin{sideways}Technique\end{sideways}} & \multicolumn{4}{l|}{\begin{sideways}Proc. Persp.\end{sideways}} & \multicolumn{4}{l|}{\begin{sideways}Application\end{sideways}} & \multicolumn{5}{l|}{\begin{sideways}Results\end{sideways}} \\\hline
    \begin{sideways}Reference\end{sideways} & \begin{sideways}Domain\end{sideways} & \begin{sideways}\end{sideways} & \begin{sideways}Technique Improv.\end{sideways} & \begin{sideways}Process Improv.\end{sideways} & \begin{sideways}Demonstration\end{sideways} & \begin{sideways}manual\end{sideways} & \begin{sideways}automated\end{sideways} & \begin{sideways}yes\end{sideways} & \begin{sideways}Real-world\end{sideways} & \begin{sideways}Synthetic\end{sideways} & \begin{sideways}Available\end{sideways} & \begin{sideways}Not available\end{sideways} & \begin{sideways}yes\end{sideways} & \begin{sideways}Imperative\end{sideways} & \begin{sideways}Declarative\end{sideways} & \begin{sideways}Hybrid\end{sideways} & \begin{sideways}Control Flow\end{sideways} & \begin{sideways}Temporal\end{sideways} & \begin{sideways}Data\end{sideways} & \begin{sideways}Resources\end{sideways} & \begin{sideways}Manual\end{sideways} & \begin{sideways}Tool-supported\end{sideways} & \begin{sideways}Automated\end{sideways} & \begin{sideways}Real-Time\end{sideways} & \begin{sideways}Qualify\end{sideways} & \begin{sideways}Quantify\end{sideways} & \begin{sideways}BD \& Comp.\end{sideways} & \begin{sideways}Loc. \& Show\end{sideways} & \begin{sideways}Expl. \& Diagn.\end{sideways}  \\\hline\hline
   \cite{accorsiExploitationProcessMining2012} & Fin & A & \TB &  &  & \TB &  & \TB &  & \TB &  & \TB & \TB & \TB &  &  & \TB & \TB & \TB & \TB &  & \TB &  &  &  & \TB & \TB & \TB & \TB \\\hline
   \cite{baumgrassConformanceCheckingRBAC2012} & Fin & A & \TB &  &  & \TB &  &  &  & \TB &  & \TB &  &  & \TB &  & \TB &  & \TB & \TB &  & \TB &  &  & \TB &  & \TB &  & \TB \\\hline
   \cite{bottrighiConformanceCheckingExecuted2012} & HC & A & \TB &  &  & \TB &  &  & \TB &  &  & \TB & \TB &  &  & \TB & \TB &  & \TB & \TB &  & \TB &  &  & \TB &  &  & \TB & \TB \\\hline
   \cite{buijsCrossOrganizationalProcessMining2012} & PS & A &  & \TB & \TB & \TB &  & \TB & \TB &  & \TB &  & \TB & \TB &  &  & \TB &  &  &  &  & \TB &  &  &  & \TB & \TB &  &  \\\hline
   \cite{chesaniAbducingComplianceIncomplete2016} & PS & A & \TB &  &  & \TB &  &  & \TB &  & \TB &  &  &  & \TB &  & \TB &  &  &  &  & \TB &  &  & \TB &  & \TB &  &  \\\hline
   \cite{chikkaConformanceCheckingMethodology2020} & HC & A &  & \TB &  & \TB &  &  & \TB &  & \TB &  & \TB & \TB &  &  & \TB &  &  &  &  &  & \TB &  &  & \TB & \TB &  &  \\\hline
   \cite{GerkeTamm2014} & ITSM & A &  & \TB & \TB & \TB &  &  & \TB &  &  & \TB & \TB & \TB &  &  & \TB &  &  &  &  & \TB &  &  &  & \TB &  & \TB &  \\\hline
   \cite{gerkeMeasuringComplianceProcesses2009} & ITSM & A & \TB &  & \TB & \TB &  &  & \TB &  &  & \TB &  & \TB &  &  & \TB &  &  &  &  & \TB &  &  &  & \TB &  &  & \TB \\\hline
   \cite{huangOnlineTreatmentCompliance2014} & HC & A & \TB & \TB &  & \TB &  & \TB & \TB &  &  & \TB & \TB &  & \TB &  & \TB & \TB & \TB & \TB &  &  & \TB & \TB & \TB &  &  &  & \TB \\\hline
   \cite{jansBusinessProcessMining2011} & Fin & A & \TB &  & \TB & \TB &  &  & \TB &  &  & \TB & \TB & \TB &  &  & \TB &  & \TB & \TB &  & \TB &  &  &  & \TB &  &  & \TB \\\hline
   \cite{jansFieldStudyUse2014} & Fin & A & \TB &  & \TB & \TB &  &  & \TB &  &  & \TB & \TB & \TB &  &  & \TB &  & \TB & \TB &  & \TB &  &  &  & \TB &  &  & \TB \\\hline
   \cite{lemosUsingProcessMining2011} & SW & A &  & \TB & \TB & \TB &  & \TB & \TB &  &  & \TB &  & \TB &  &  & \TB &  &  &  &  & \TB &  &  &  & \TB & \TB & \TB & \TB \\\hline
   \cite{lenkowiczAssessingConformityClinical2018} & HC & A &  & \TB & \TB & \TB &  & \TB & \TB &  &  & \TB & \TB & \TB &  &  & \TB & \TB & \TB &  &  & \TB &  &  &  & \TB &  & \TB & \TB \\\hline
   \cite{liangIdentificationHumanOperation2021} & Nuc & A & \TB &  &  & \TB &  &  & \TB &  &  & \TB &  & \TB &  &  & \TB &  &  &  &  & \TB &  &  &  & \TB &  &  &  \\\hline
   \cite{piovesanTemporalConformanceAnalysis2018} & HC & A & \TB &  &  & \TB &  &  & \TB &  &  & \TB &  &  & \TB &  & \TB & \TB & \TB & \TB &  & \TB &  &  & \TB &  &  & \TB & \TB \\\hline
   \cite{placidiProcessMiningOptimize2021} & HC & A &  & \TB & \TB & \TB &  & \TB & \TB &  &  & \TB & \TB & \TB &  &  & \TB & \TB & \TB &  &  & \TB &  &  &  & \TB & \TB & \TB &  \\\hline
   \cite{premchaiswadiProcessModelingBottleneck2015} & Sci & A &  & \TB &  & \TB &  &  & \TB &  &  & \TB &  & \TB &  &  & \TB &  &  &  &  & \TB &  &  &  & \TB &  & \TB & \TB \\\hline
   \cite{rinnerProcessMiningConformance2018} & HC & A &  &  & \TB & \TB &  &  & \TB &  &  & \TB & \TB & \TB &  &  & \TB & \TB &  &  &  & \TB &  &  &  & \TB & \TB & \TB & \TB \\\hline
   \cite{rovaniDeclarativeProcessMining2015} & HC & A & \TB & \TB & \TB & \TB &  &  & \TB &  &  & \TB & \TB &  & \TB &  & \TB &  &  &  &  & \TB &  &  &  & \TB & \TB & \TB & \TB \\\hline
   \cite{sangilHeuristicsBasedProcessMining2020} & PS & A &  & \TB & \TB & \TB &  &  & \TB &  & \TB &  & \TB & \TB &  &  & \TB & \TB &  &  &  & \TB &  &  &  & \TB & \TB &  &  \\\hline
   \cite{sarnoHybridAssociationRule2015} & Fin & A & \TB &  &  & \TB &  & \TB &  & \TB &  & \TB & \TB &  &  & \TB & \TB & \TB & \TB & \TB &  & \TB &  &  &  & \TB & \TB &  &  \\\hline
   \cite{satoConformanceCheckingDifferent2020} & HC & A & \TB &  & \TB & \TB &  & \TB & \TB &  &  & \TB & \TB & \TB &  &  & \TB &  &  &  &  & \TB &  &  &  & \TB & \TB &  &  \\\hline
   \cite{spiottaTemporalConformanceAnalysis2017} & HC & A & \TB &  &  & \TB &  &  & \TB &  &  & \TB & \TB &  & \TB &  & \TB & \TB & \TB & \TB &  & \TB &  &  & \TB &  &  &  & \TB \\\hline
   \cite{stertzRoleTimeData2021} & Man & A & \TB & \TB & \TB & \TB &  & \TB &  & \TB & \TB &  & \TB & \TB &  &  & \TB & \TB & \TB & \TB &  &  & \TB & \TB &  & \TB &  &  & \TB \\\hline
   \cite{taghiabadiComplianceCheckingDataAware2014} & HC,Fin & A & \TB &  &  & \TB &  &  & \TB & & \TB & \TB &  & \TB &  &  & \TB &  & \TB & \TB &  & \TB &  &  &  & \TB &  &  & \TB \\\hline
   \cite{xuModelingClinicalActivities2020} & HC & A & \TB & \TB & \TB & \TB &  & \TB & \TB &  &  & \TB & \TB &  & \TB &  & \TB & \TB & \TB &  &  & \TB &  &  &  & \TB & \TB &  & \TB \\\hline
   \cite{xuApplyingClinicalGuidelines2020} & HC & A & \TB &  & \TB & \TB &  & \TB & \TB &  &  & \TB & \TB &  & \TB &  & \TB & \TB & \TB & \TB &  & \TB &  &  &  & \TB &  & \TB & \TB \\\hline
   \cite{zamanInnovativeOnlineProcess2019} & DP & A & \TB &  &  & \TB &  &  &  & \TB &  & \TB &  & \TB &  &  & \TB & \TB & \TB &  &  & \TB &  &  &  & \TB &  & \TB & \TB \\\hline
   \cite{zamanEnablingGDPRCompliance2020} & DP & A & \TB &  &  & \TB &  &  &  & \TB &  & \TB &  & \TB &  &  & \TB & \TB & \TB &  &  & \TB &  &  &  & \TB &  & \TB &  \\\hline
   \cite{beneventoProcessModeling2023} & HC & A &  & \TB & \TB & \TB &  & \TB & \TB &  &  & \TB & \TB & \TB &  &  & \TB &  &  &  &  & \TB &  &  &  & \TB &  &  & \TB \\\hline
   \cite{grugerVerifyingGuideline2022} & HC & A & \TB &  & \TB & \TB &  &  & \TB &  &  & \TB & \TB & \TB &  &  & \TB &  & \TB &  &  & \TB &  &  &  & \TB &  &  &  \\\hline
   \cite{grugerWeightedViolations2023} & HC & A & \TB &  &  & \TB &  &  & \TB &  &  & \TB & \TB &  & \TB &  & \TB &  & \TB &  &  & \TB &  &  &  & \TB & \TB & \TB & \TB \\\hline
   \cite{savinoProcessMining2023} & HC & A &  & \TB & \TB & \TB &  &  & \TB &  &  & \TB & \TB &  & \TB &  & \TB &  & \TB &  &  & \TB &  &  &  & \TB &  & \TB & \TB \\\hline
   \cite{grugerDeclarativeGuideline2023} & HC & A & \TB &  & \TB & \TB &  &  & \TB &  &  & \TB & \TB &  & \TB &  & \TB &  & \TB &  &  & \TB &  &  &  & \TB &  &  &  \\\hline
   \cite{ramezani2012WhereDidI} & HC & A & \TB & & & \TB & & & \TB & & & \TB & & \TB & & & \TB & & \TB & & & \TB & & & & \TB & & \TB & \TB \\\hline
   \cite{breitmayerPermissionAnalysisObjectCentric2024} & Edu & A & \TB & & \TB & \TB & & & \TB & & \TB & & & & \TB & & \TB & & \TB & \TB & & \TB & & & \TB & & & & \TB \\\hline
   \cite{savinoErrorCorrectingMethodologyEvaluating2024} & HC & A & \TB & & & \TB & & & \TB & & & \TB & \TB & \TB & & & \TB & & \TB & & & \TB & & & \TB & \TB & \TB & \TB & \TB \\\hline
   \cite{zhangReorderedFuzzyConformance2024} & HC & A & \TB & & & \TB & & & \TB & & & \TB & \TB & \TB & & & \TB & & & & & \TB & & & & \TB & \TB & & \TB \\\hline
   \cite{salnitriSecuritybyDesignIdentificationSecurityCritical2018} & Fin & A & \TB & & & \TB & & & \TB & & \TB & & & \TB & & & \TB & & & \TB & & \TB & & & & \TB & & \TB & \TB \\\hline
   \cite{wangRegulatorySupervisionComputational2018a} & Sup & A & & \TB & \TB & \TB & & & \TB & & & \TB & \TB & \TB & & & \TB & & & & & \TB & & & & & & \TB & \TB \\\hline
    \cite{wangAnalyzingTransactionCodes2018a} & Man & A & & \TB & & \TB & & \TB & \TB & & & \TB & & \TB & & & \TB & & & & & \TB & & & & & & \TB & \TB \\\hline\hline

   \cite{knupleschAFramework2017} & HR & B & \TB &  & \TB & \TB &  &  & \TB & \TB &  & \TB &  &  & \TB &  & \TB & \TB & \TB & \TB &  & \TB &  &  & \TB & \TB & \TB & \TB & \\\hline
   \cite{lyComplianceMonitoringBusiness2015} & HC & B &  &  & \TB & \TB &  &  & \TB &  & \TB &  & \TB &  & \TB &  & \TB & \TB & \TB &  &  & \TB &  &  &  & \TB & \TB & \TB & \\\hline
   \cite{vanbeestCrossInstanceRegulatoryCompliance2023} & Sup & B & \TB &  &  & \TB &  &  & \TB &  & \TB &  &  &  & \TB &  & \TB & \TB & \TB & \TB &  & \TB &  &  &  & \TB &  & \TB & \\\hline
   \cite{araujoAMethod2010} & Rent & B & \TB &  &  & \TB &  &  &  & \TB &  & \TB &  &  & \TB &  & \TB & \TB &  & \TB &  & \TB &  &  &  &  &  & \TB & \\\hline
   \cite{jiang2014RegulatoryCompliance} & Trade & B &  & \TB &  & \TB &  &  &  & \TB &  & \TB &  &  & \TB &  & \TB & \TB & \TB &  &  & \TB &  &  & \TB &  &  & \TB & \TB\\\hline
   \cite{jiangComplianceChecking2015} & Trade & B &  & \TB &  & \TB &  & \TB &  & \TB &  & \TB & \TB &  & \TB &  & \TB & \TB & \TB & \TB &  & \TB &  &  & &  &  & \TB & \\\hline
   \cite{weidlichEventBasedMonitoring2011} & CS & B & \TB &  &  & \TB &  &  & \TB &  &  & \TB & \TB & \TB &  &  & \TB &  & \TB &  &  & \TB &  &  &  & \TB &  &  & \TB\\\hline
   \cite{hashmiNormativeRequirementsBusiness2014} & PS & B & \TB &  &  & \TB &  & \TB & \TB &  &  & \TB &  &  & \TB &  & \TB &  & \TB & \TB &  & \TB &  &  &  & \TB &  &  & \TB \\\hline
   \cite{WOS:000540648600033} & DP & B & & & \TB & \TB & & & \TB & & \TB & & \TB & & \TB & & \TB & \TB & \TB & & & \TB & & & & \TB & & & \TB \\\hline
   \cite{bukhshComplianceCheckingShipment2017} & Trade & B & & & \TB & \TB & & & & \TB & & \TB & & & \TB & & \TB & & \TB & & & \TB & & & & \TB & & & \TB \\\hline
   \cite{caronBusinessRulePatterns2013} & Fin,HC & B & & & \TB & \TB & & & \TB & & & \TB & & & \TB & & \TB & & \TB & \TB & & \TB & & & & \TB & & & \TB \\\hline
   \cite{gongBpCMonRulebasedMonitoring2017} & HC & B & \TB & & & \TB & & & \TB & & \TB & & & & \TB & & \TB & \TB & \TB & \TB & & \TB & & & & \TB & & & \\\hline
   \cite{gonzalezComplianceRequirements2021} & PS & B & \TB & & & \TB & & & \TB & & & \TB & & & \TB & & \TB & \TB & \TB & \TB & & \TB & & & \TB & \TB & & & \\\hline 
   \cite{hoangChallengesDiscoveringPatient2022} & HC & B & & & \TB & \TB & & & \TB & & & \TB & \TB & \TB & & & \TB & & & & & \TB & & & & \TB & & \TB & \\\hline
   \cite{lyMonitoringBusinessProcess2011} & Fin & B & \TB & & & \TB & & & & \TB & & \TB & & & \TB & & \TB & \TB & \TB & & & \TB & & & & & & \TB & \TB \\\hline
   \cite{middletonFrameworkContinuousCompliance2009} & HC & B & \TB & & & \TB & & & \TB & & & \TB & & & \TB & & \TB & \TB & \TB & & & & \TB & \TB & & & & \TB & \\\hline
   \cite{WOS:000424457700006} & Fin & B & \TB & & \TB & \TB & & & \TB & & & \TB & \TB & & \TB & & \TB & \TB & \TB & & & \TB & & & \TB & & & & \TB \\\hline

\end{tabularx}
\label{tab:coding}
\end{table}

For the \emph{preparation} phase, we capture the process \emph{domain}, the \emph{goal} with which a compliance monitoring technique is applied (to \emph{improve} compliance monitoring techniques, to improve the business \emph{process compliance}, or to \emph{demonstrate} compliance monitoring techniques), as well as whether the prescriptive model is \emph{created} manually or automatically, and whether it is explicitly \emph{validated} through e.g., process experts or stakeholders. We also capture whether the process execution data in the form of event logs has synthetic or real-world \emph{origins}, and whether it is publicly \emph{available} or not.

For the \emph{action} phase, we identify whether \emph{expert knowledge} is necessary for applying compliance monitoring techniques, whether imperative, declarative, or hybrid \emph{techniques}\footnote{This we determine primarily based on the type of prescriptive model --- i.e., declarative, imperative, or hybrid --- since the exact technique for regulatory compliance monitoring was not made transparent in all studies we investigated. Nonetheless, the way in which compliance constraints are formalized appears to be a good indicator of the type of technique applied to them.} are used (see~\cite{carmona2018conformance,de2015alignment}), which \emph{process perspectives} are assessed (control flow, temporal, data, or resource, see~\cite{russellWorkflowPatternsDefinitive2015}), and whether the compliance monitoring techniques are \emph{applied} manually, in a tool-supported manner, automatically, or in real-time (i.e., \emph{online}). In contrast to~\cite{lyComplianceMonitoringBusiness2015}, we also explicitly include the control flow process perspective. However, we do so on a higher level of abstraction than the control flow requirements formulated by~\cite{rinderle-maPredictiveComplianceMonitoring2023}, which differentiate control flow constraints between existence, absence, and ordering constraints. Subsuming these under the control flow perspective, as in~\cite{dunzerConformanceCheckingStateoftheart2019}, has been done to describe the application of regulatory compliance monitoring techniques more broadly.

For the \emph{reflection} phase, we capture the different \emph{types of results} and their representation (i.e., whether compliance is \emph{qualified}, \emph{quantified}, \emph{broken down and compared}, or \emph{explained and diagnosed}, following a classification of representations identified from current process mining tools~\cite{rehse2022process}). Here, we focus on understanding why and how results of regulatory compliance monitoring are presented, as opposed to determining whether existing approaches are \emph{able} to provide specific insights to users~\cite{lyComplianceMonitoringBusiness2015,rinderle-maPredictiveComplianceMonitoring2023}. Doing so, we find that qualifying the overall degree of regulatory compliance is also relevant in concrete applications, something not explicitly covered by related work~\cite{lyComplianceMonitoringBusiness2015,rinderle-maPredictiveComplianceMonitoring2023}.

For Search A, the screening, data extraction, and synthesis steps were performed by both authors and a research assistant. The initial papers found in Search A were distributed among them for further analysis as described above, and the results of that analysis were combined in a discussion. During this discussion, unclarities and disagreements were resolved. For Search B, the main author and another research assistant performed the screening, data extraction, and synthesis steps in a similar manner, while re-using the coding scheme developed during Search A. Again, they combined the results of their analyses in a discussion, and resolved unclarities and disagreements. Finally, the main author performed an assessment of the entire result of Search A and B, and discussed final changes to the analysis among the author team, until a consensus was reached.
In Table~\ref{tab:categories_explained}, we provide a detailed description of the categories and characteristics we extracted in Table~\ref{tab:coding}.
In the following, we analyse the resulting synthesis along its dimensions, thereby \emph{writing the review}.

\section{Results}
\label{sec:results-analysis}

After inductive coding and synthesizing the 58 (41 of Search A, 17 of Search B) relevant studies, we analyse the resulting data, in order to characterize their compliance monitoring mechanisms and their usages. This is done along the three phases of conformance checking projects~\cite{carmona2018conformance}, being \emph{preparation} (Sec.~\ref{sec:results-prep}), \emph{action} (Sec.~\ref{sec:results-action}), and \emph{reflection} (Sec,~\ref{sec:results-reflection}). We also describe commonalities and differences between the uses found in the two searches.
Generally, we can observe, as shown in Fig.~\ref{fig:study-distribution}, that we find far fewer studies that position themselves as compliance checking than those framed as conformance checking studies. We also note that while conformance checking studies were published quite early, they appear to peak slightly later than compliance checking studies, which might hint at a shift in research interest or a change in the positioning of the studies; similarly, compliance checking studies seem to tail off after having reached a peak in 2017. %

\begin{figure}[h!]
    \centering
    \begin{tikzpicture}
    \begin{axis}[
        height=5cm,
        width=\textwidth,
        x tick label style={
            /pgf/number format/1000 sep=},
        ylabel=Studies,
        ymax = 7,
        xlabel=Year,
        legend style={at={(0.25,1.25)},
            anchor=north,legend columns=-1},
        ybar,
        bar width=4pt,
        xtick=data,
        xticklabel style={rotate=90},
        nodes near coords,
        nodes near coords align={vertical}, 
    ]
        \addplot 
    	coordinates {(2009,2) (2010,0) (2011,2) (2012,5) (2013,0) (2014,3) (2015,3) (2016,1) (2017,1) (2018,6) (2019,1) (2020,6) (2021,3) (2022,1) (2023,4) (2024,3)};
    \addplot 
    	coordinates {(2009,1) (2010,1) (2011,2) (2012,0) (2013,1) (2014,2) (2015,2) (2016,0) (2017,3) (2018,1) (2019,1) (2020,0) (2021,1) (2022,1) (2023,1) (2024,0)};
     
        \legend{Conformance,Compliance}
    \end{axis}
    \end{tikzpicture} 
    \caption{Distribution of studies found for Search A (\emph{conformance}) and Search B (\emph{compliance}) per year}
    \label{fig:study-distribution}
\end{figure}

\subsection{Preparation}
\label{sec:results-prep}

\paragraph{Domain} First, we analyse the domains in which conformance checking and compliance checking techniques are applied for regulatory compliance monitoring. Figure~\ref{fig:domain-distribution} shows an overview of the domains we identified.

\begin{figure}[ht!]
	\centering
	\begin{tikzpicture}
		\begin{axis}[
			height=5cm,
			width=\textwidth,
			ylabel=Studies,
			ymax=25,
			xlabel=Domain,
			symbolic x coords={CS, DP, Edu, Fin, HC, HR, ITSM, Man, Nuc, PS, Rent, Sci, Sup, SW, Trade},
			legend style={at={(0.25,1.25)},
				anchor=north,legend columns=-1},
			ybar,
			bar width=4pt,
			xtick=data,
			xticklabel style={rotate=90},
			nodes near coords,
			nodes near coords align={vertical}, 
			]
			\addplot 
			coordinates {(CS,0) (DP,2) (Edu,1) (Fin,7) (HC,21) (HR,0) (ITSM,2) (Man,2) (Nuc,1) (PS,3) (Rent,0) (Sci,1) (Sup,1) (SW,1) (Trade,0)};
			\addplot 
			coordinates {(CS,1) (DP,1) (Edu,0) (Fin,3) (HC,5) (HR,1) (ITSM,0) (Man,0) (Nuc,0) (PS,2) (Rent,1) (Sci,0) (Sup,1) (SW,0) (Trade,3)};
			
			\legend{Conformance,Compliance}
		\end{axis}
	\end{tikzpicture} 
	\caption{Distribution of domains found in the studies identified with Search A (\emph{conformance}) and Search B (\emph{compliance})}
	\label{fig:domain-distribution}
\end{figure}

The majority of \emph{conformance} checking studies focusses on the area of healthcare (21 studies), with finance being second (seven studies). Other areas, such as public services (three studies) or data processing (two studies), are also present.
The prevalence of healthcare and finance seems to indicate that regulations in these areas are well-suited to be assessed with conformance checking. %

The set of \emph{compliance} checking studies, on the other hand, shows a broader distribution of domains (including one study in a healthcare setting): From healthcare, human resources, rent, public services to trade, supply chains and customer service, seven different domains are covered. This seems to suggest a broad applicability of compliance checking in many areas, although we observe a slight focus on healthcare-related studies (five times), trade and supply chain (four times), as well as finance and public services (three times, two times).

\paragraph{Goals} Second, we consider the goals of the studies we analysed. An overview of our findings is provided in Fig.~\ref{fig:goals-distribution}.

\begin{figure}[ht!]
	\centering
	\begin{tikzpicture}
		\begin{axis}[
			height=5cm,
			x=15ex,
			width=8cm,
			ylabel=Studies,
			enlarge x limits=0.3,
			xlabel=Goal,
			ymax = 33,
			symbolic x coords={,{Technique\\Improvement}, {Process\\Improvement}, {Demonstration},},
			legend style={at={(0.25,1.25)},
				anchor=north,legend columns=-1},
			ybar,
			bar width=6pt,
			xtick=data,
			xticklabel style={align=center},
			nodes near coords,
			nodes near coords align={vertical}, 
			]
			\addplot 
			coordinates {({Technique\\Improvement},28) ({Process\\Improvement},16) ({Demonstration},21)};
			\addplot 
			coordinates {({Technique\\Improvement},10) ({Process\\Improvement},2) ({Demonstration},7)};
			
			\legend{Conformance,Compliance}
		\end{axis}
	\end{tikzpicture} 
	\caption{Distribution of goals found in the studies identified with Search A (\emph{conformance}) and Search B (\emph{compliance})}
	\label{fig:goals-distribution}
\end{figure}

Regarding the explicitly stated goals of the studies for applying \emph{conformance} checking techniques, we see that those in the area of healthcare mainly aim to improve the conformance checking technique and its application (15 of 21) or to demonstrate the applicability of conformance checking (eleven of 21), whereas only around a third of the studies making use of healthcare-related regulations aim to improve the process itself. Similarly, no finance-related contribution aims to provide actual process improvement, but all of them aim to improve conformance checking techniques, and two out of seven explicitly intend to demonstrate them. 
This tendency is noticeably present across all domains we identified, with more than two thirds (28) aiming to improve techniques, more than half aiming to demonstrate them (21), and less than half of the studies (16) aiming to investigate actual process improvement.

For \emph{compliance} checking applications, the situation is slightly: more than half (i.e., ten of 17) of the studies aim to improve compliance checking techniques and their application, only two of 17 studies aim to improve the business process itself, and little over two fifths (i.e., seven of 17 studies) aim to demonstrate the application of compliance checking. Notably, all three studies that explicitly set out to improve process compliance are situated in the area of trade and supply chains.

\paragraph{Prescriptive Model Creation} Third, we consider how prescriptive models are created in the relevant studies. In all \emph{conformance} checking and \emph{compliance} checking studies, we see that no contribution out of the 58 identified studies automatically creates them based on the relevant regulations.
In fact, all studies rely on manual input for creation of the required prescriptive model. Notably, the exact operationalization that led to the model is largely not described in detail, and instead, only descriptions or visualizations of them are provided.
For example, in~\cite{accorsiExploitationProcessMining2012}, a BPMN model is created manually with and validated by process experts (in this case, bank managers), and further requirements are provided informally by bank managers and auditors, and manually translated into LTL formulae by the authors.

\paragraph{Prescriptive Model Validation} Additionally, we note that in less than one third of the \emph{conformance} checking studies (13 of 41), the prescriptive model is explicitly validated by the authors in conjunction with stakeholders or process experts.\footnote{The remaining studies did not mention whether the prescriptive model was evaluated or not. Hence, we only capture whether an explicit evaluation took place.} Interestingly, around half (seven of 13) of those studies in which a validation is reported are in the healthcare domain.
We assume that this may be because of the need for expert involvement in healthcare settings to even operationalize the treatment guidelines, so that validation is a comparatively trivial addition.

In contrast to this, only two of the 17 \emph{compliance} checking studies explicitly mention a validation of the prescriptive model that has been used.\footnote{As above, we only noted explicit evaluations, since the remaining seven studies did not mention whether an evaluation took place or not.} This may be linked to the prevalent technique improvement goal we see in compliance checking studies: in order to improve compliance checking techniques, the prescriptive models may not need to be evaluated since an evaluated prescriptive model is less relevant here than for demonstrating techniques or improving actual process compliance.

\paragraph{Data Origin and Data Availability}
For the final aspect of the preparation phase, we look at the origin and availability of the process event data used in the analysed studies. In Fig.~\ref{fig:data_origin_availability}, we provide an overview of our findings.

\begin{figure}[ht!]
	\begin{subfigure}{.49\textwidth}
		\centering
		\begin{tikzpicture}
			\begin{axis}[
				height=5cm,
				x=3cm, %
				enlarge x limits=0.3,
				width=\textwidth,
				ylabel=Studies,
				xlabel=Origin,
				ymax = 40,
				symbolic x coords={,Real-world,Synthetic,},
				legend style={at={(0.5,1.25)},
					anchor=north,legend columns=-1},
				ybar,
				bar width=6pt,
				xtick=data,
				nodes near coords,
				nodes near coords align={vertical}, 
				]
				\addplot 
				coordinates {(Real-world,35) (Synthetic,6)};
				\addplot 
				coordinates {(Real-world,12) (Synthetic,6)};
				
				\legend{Conformance,Compliance}
			\end{axis}
		\end{tikzpicture} 
		\caption{Data Origin}
		\label{fig:data_origin}
	\end{subfigure}
	\begin{subfigure}{.49\textwidth}
		\centering
		\begin{tikzpicture}
			\begin{axis}[
				height=5cm,
				x=3cm, %
				enlarge x limits=0.3,
				width=\textwidth,
				ylabel=Studies,
				xlabel=Origin,
				ymax = 40,
				symbolic x coords={,Available, {Not Available},},
				ybar,
				bar width=6pt,
				xtick=data,
				nodes near coords,
				nodes near coords align={vertical}, 
				]
				\addplot 
				coordinates {(Available,8) ({Not Available},34)};
				\addplot 
				coordinates {(Available,4) ({Not Available},13)};
				
			\end{axis}
		\end{tikzpicture} 
		\caption{Data Availability}
		\label{fig:data_availability}
	\end{subfigure}
	\caption{Data Origin and data availability for the studies identified with Search A (\emph{conformance}) and Search B (\emph{compliance})}
	\label{fig:data_origin_availability}
\end{figure}

As to the \emph{conformance} checking studies, 35 of 41 conformance checking studies use real-world data, although only in seven cases this data is available. Usually, available real-world process event data stems from the \emph{Business Process Intelligence Challenge} (BPIC) (e.g.,~\cite{taghiabadiComplianceCheckingDataAware2014}), but also from data provided by a governmental data sharing portal~\cite{sangilHeuristicsBasedProcessMining2020} or a publicly available clinical database~\cite{chikkaConformanceCheckingMethodology2020}. 21 of 35 applications with real-world data are in the healthcare domain, which is also the most represented domain in the conformance checking studies. On the other hand, only six of 41 studies use purely synthetic data, which is made available in one case. We therefore observe a low overall availability of process event data in the conformance checking space but a high degree of real-world data. 

In the \emph{compliance} checking studies, we see real-world process event data being used in twelve of 17 studies, and synthetic data in six of 17 studies. However, only in four of the real-world studies, this data is actually available -- notably, three of them employ BPIC event logs which have been made available for research purposes beforehand~\cite{lyComplianceMonitoringBusiness2015,vanbeestCrossInstanceRegulatoryCompliance2023,gongBpCMonRulebasedMonitoring2017}.

\subsection{Action}
\label{sec:results-action}

\paragraph{Technique} The \emph{conformance} techniques used in the investigated contributions, are largely imperative, with less than a third (i.e., eleven of 41) declarative and only two hybrid techniques. The largest share of declarative techniques is present in the healthcare domain (eight of eleven), underlining that the nature of clinical guidelines may be especially suited for declarative techniques. For example,~\cite{bottrighiConformanceCheckingExecuted2012} translates a clinical guideline into a formal representation based on logical programming that combines declarative and imperative aspects. Based on a Prolog implementation of the \emph{Event Calculus}, they can assess the regulatory compliance of recorded process executions w.r.t. a clinical guideline. As another example,~\cite{sangilHeuristicsBasedProcessMining2020} uses ProM to compare the activities and timings of public procurement data provided by a government's open data portal with the activity orderings and timings prescribed by relevant regulation.

For \emph{compliance} checking techniques, we observe almost exclusively declarative applications (15 out of 17), regardless of domain. For example,~\cite{bukhshComplianceCheckingShipment2017} use ProM's LTL checker to assess the regulatory compliance of a company's trade and customs procedures, for which they manually create formal regulatory constraints from relevant statements of the regulations. The prescriptive model is imperative in nature only in two studies, for example in~\cite{weidlichEventBasedMonitoring2011} where it serves to derive event queries for use with CEP. The overall prevalence of declarative applications may be linked to mechanisms used for regulatory compliance monitoring, such as CEP, where instead of alignments with an imperative process model, done so in imperative conformance checking techniques~\cite{carmona2018conformance}, \emph{standing queries} are evaluated against observed process event data~\cite{weidlichEventBasedMonitoring2011} to determine compliance or non-compliance.

\paragraph{Process Perspective} As to the process perspective, we provide an overview of our findings in Fig.~\ref{fig:process_perspective}.

\begin{figure}[ht!]
	\begin{subfigure}{.48\textwidth}
		\centering
		\begin{tikzpicture}
			\begin{axis}[
				width=0.8\textwidth,
				height=5cm,
				xlabel=Technique,
				ylabel=Process Perspective,
				colormap name=myBlues, %
				point meta min=1, %
				point meta max=30, %
				xtick=data,
				ytick=data,
				xticklabel style={rotate=45, anchor=east},
				symbolic x coords={Imperative, Declarative, Hybrid},
				symbolic y coords={{Control Flow}, Temporal, Data, Resource},
				]
				\addplot[
				matrix plot*,
				mesh/cols=3,
				nodes near coords,
				nodes near coords align={center}, %
				point meta=explicit, %
				] 
				table [meta=z] {
					x y z
					Imperative {Control Flow} 27
					Declarative {Control Flow} 11
					Hybrid {Control Flow} 2
					Imperative Temporal 8
					Declarative Temporal 5
					Hybrid Temporal 1
					Imperative Data 13
					Declarative Data 9
					Hybrid Data 2
					Imperative Resource 6
					Declarative Resource 6
					Hybrid Resource 2
				};
			\end{axis}
		\end{tikzpicture}
		\caption{Conformance studies}
	\end{subfigure}
	\begin{subfigure}{.48\textwidth}
		\begin{tikzpicture}
			\begin{axis}[
				width=0.8\textwidth,
				height=5cm,
				xlabel=Technique,
				ylabel=Process Perspective,
				colormap name=myReds, %
				point meta min=0, %
				point meta max=20, %
				xtick=data,
				ytick=data,
				xticklabel style={rotate=45, anchor=east},
				symbolic x coords={Imperative, Declarative, Hybrid},
				symbolic y coords={{Control Flow}, Temporal, Data, Resource},
				]
				\addplot[
				matrix plot*,
				mesh/cols=3,
				nodes near coords,
				nodes near coords align={center}, %
				point meta=explicit, %
				] 
				table [meta=z] {
					x y z
					Imperative {Control Flow} 2
					Declarative {Control Flow} 15
					Hybrid {Control Flow} 0
					Imperative Temporal 0
					Declarative Temporal 12
					Hybrid Temporal 0
					Imperative Data 1
					Declarative Data 14
					Hybrid Data 0
					Imperative Resource 0
					Declarative Resource 8
					Hybrid Resource 0
				};
			\end{axis}
		\end{tikzpicture}
		\caption{Compliance studies}
	\end{subfigure}
	\caption{Techniques and process perspectives of the studies identified with Search A (\emph{conformance}) and Search B (\emph{compliance})}
	\label{fig:process_perspective}
\end{figure}

We observe that control-flow constraints are always assessed in the 41 studies that apply \emph{conformance} checking. All other perspectives, being temporal (14 times), data (24 times), and resource (14 times), are used in various combinations in around half of the applications, with a prevalence of data constraints. Notable is that all hybrid applications assess both data and resource perspectives. Hence, relevant process perspectives seem to be important in the choice of technique.

\emph{Compliance} checking techniques, on the other hand, appear more multifaceted --- almost always more than two process perspectives considered: all 17 studies assess control flow aspects, while twelve of 17 times temporal, and 15 of 17 times data aspects are assessed; the resource perspective is used eight of 17 times. Notably, two of the five compliance checking studies where the temporal process perspective is not considered are also the two imperative compliance checking studies we found in our search~\cite{weidlichEventBasedMonitoring2011,hoangChallengesDiscoveringPatient2022}. This appears to be related to the imperative nature of the approach's technique, where temporal aspects were not formalized, and thus, as in the case of conformance checking, the process perspectives seem related to the choice of technique.

\paragraph{Application} We identify that the majority of \emph{conformance} checking applications (38 of 41) are conducted with tools, but not in an automated or real-time fashion. More concretely, we only note three applications that are automated and two that are real-time. 
One of these real-time studies proposes an automated approach, where clinical events created during patient treatment of unstable clinical angina are pre-processed using i.a. NLP techniques, constructed into traces, and checked against compliance rules describing the treatment process. Potential violations are sent directly as feedback to the clinical information system, allowing clinicians to react and remedy or accept compliance violations~\cite{huangOnlineTreatmentCompliance2014}. 
Finally, all other applications rely on tool support for conformance checking; for example, in~\cite{xuApplyingClinicalGuidelines2020} ProM is used for checking conformance to a Declare-based prescriptive model.
From this, we conclude that existing conformance checking tools, especially ProM, have been disseminated widely and reached acceptance in the scientific community. Notably, automated/real-time conformance checking applications are limited to the healthcare and manufacturing domain, potentially due to the presence of information systems in hospitals and production chains allowing for detailed event log recording and automatic application of conformance checking techniques.

For the \emph{compliance} checking studies, we identify tool-supported uses in all but one of 17 cases. Concretely, we observe multiple prototypical implementations~\cite{weidlichEventBasedMonitoring2011,knupleschAFramework2017,vanbeestCrossInstanceRegulatoryCompliance2023,middletonFrameworkContinuousCompliance2009,gongBpCMonRulebasedMonitoring2017,WOS:000424457700006}, which usually take an event log as input that is replayed against the constraints of a provided prescriptive model. We also see ProM plugins being used~\cite{lyComplianceMonitoringBusiness2015,gonzalezComplianceRequirements2021,bukhshComplianceCheckingShipment2017}, and tools such as the SeaFlows\footnote{See \url{https://www.uni-ulm.de/in/iui-dbis/forschung/abgeschlossene-projekte/seaflows/} [Accessed: 18/11/2024]} compliance checking framework~\cite{lyComplianceMonitoringBusiness2015,lyMonitoringBusinessProcess2011} or the Regorous tool set\footnote{See \url{https://research.csiro.au/data61/regorous/} [Accessed: 11/11/2024]}~\cite{hashmiNormativeRequirementsBusiness2014} used for compliance checking. Notably, only in one instance an automated, real-time technique is applied, where compliance is monitored continuously and alerts are triggered in real-time~\cite{middletonFrameworkContinuousCompliance2009}.

\paragraph{Expert Knowledge} For the final aspect of the action phase, we consider the need for expert knowledge in uses of regulatory compliance monitoring. Figure~\ref{fig:expert_knowledge_domain} displays an overview of our findings.

\begin{figure}[ht!]
	\centering
	\begin{tikzpicture}
		\begin{axis}[
			height=5cm,
			width=\textwidth,
			ylabel=\% of Studies,
			ymax=115,
			xlabel=Domain,
			symbolic x coords={CS, DP, Edu, Fin, HC, HR, ITSM, Man, Nuc, PS, Rent, Sci, Sup, SW, Trade},
			legend style={at={(0.25,1.25)},
				anchor=north,legend columns=-1},
			ybar,
			bar width=4pt,
			xtick=data,
			xticklabel style={rotate=90},
			nodes near coords,
			nodes near coords align={vertical},
			every node near coord/.append style={font=\footnotesize}
			]
			\addplot 
			coordinates {(CS,0) (DP,0) (Edu,0) (Fin,57.14) (HC,85.71) (HR,0) (ITSM,50) (Man,50) (Nuc,0) (PS,66.67) (Rent,0) (Sci,0) (Sup,100) (SW,0) (Trade,0)};
			\addplot 
			coordinates {(CS,100) (DP,100) (Edu,0) (Fin,33.33) (HC,40) (HR,0) (ITSM,0) (Man,0) (Nuc,0) (PS,0) (Rent,0) (Sci,0) (Sup,0) (SW,0) (Trade,33.33)};
			
			\legend{Conformance,Compliance}
		\end{axis}
	\end{tikzpicture} 
	\caption{Share of explicitly required expert knowledge per domain for the studies identified with Search A (\emph{conformance}) and Search B (\emph{compliance})}
	\label{fig:expert_knowledge_domain}
\end{figure}

Of all the \emph{conformance} checking contributions, around two thirds (27 of 41) explicitly reference the need for expert knowledge, either in operationalizing the regulations into a prescriptive model for conformance checking, in the application of conformance checking, or in interpreting the results.\footnote{Similar to the model validation, the remaining studies made no explicit reference to such a requirement; therefore we only differentiate between explicit needs for expert knowledge and no reference to such a need.} %
For example,~\cite{xuApplyingClinicalGuidelines2020} relies on clinical experts for interpreting the conformance checking results, and~\cite{sangilHeuristicsBasedProcessMining2020} conducts a validation interview with procuring entities to confirm and contextualize the respective findings.
Further, how expert knowledge concretely contributes is often not mentioned in detail, but instead, only the resulting prescriptive model or interpretation is presented (e.g., in~\cite{satoConformanceCheckingDifferent2020}, the use of expert knowledge in the form of a medical specialist for the definition of the case study is mentioned, but the exact input of the expert and to what end remain not known).
Notably, we see that 18 out of 21 applications in healthcare, two of three applications in public services and four out of seven applications in finance explicitly require expert knowledge throughout the application, a higher ratio than in any other application domain with at least two studies that we identified. This might be explained by the difficulty of operationalizing regulations and deriving prescriptive models for conformance checking in this domain compared to others. A potential reason might lie in the language and wording specific to these domains, or the implicit domain knowledge required for interpreting the regulations correctly.

As to the \emph{compliance} checking studies, six of 17 (i.e., slightly more than a third) explicitly mention a need for expert knowledge. For example,~\cite{weidlichEventBasedMonitoring2011} mentions the need for a domain expert to determine appropriate timeout settings in the application of their approach.\footnote{As above, we exclusively identified explicitly mentioned needs for expert knowledge.} 
This lower share of studies with explicitly required expert knowledge might also tie back to the primary purpose of improving compliance checking techniques, which, compared to improving actual process compliance, would require less expert aid in operationalizing rules for regulatory compliance monitoring.

\subsection{Reflection}
\label{sec:results-reflection}

Finally, we analyse the results and their visualization. An overview of our findings is provided in Fig.~\ref{fig:analysis_results}.

\begin{figure}[!ht]
	\begin{subfigure}{.48\textwidth}
		\centering
		\begin{tikzpicture}
			\begin{axis}[
				width=0.8\textwidth,
				height=5cm,
				xlabel=Technique,
				ylabel=Results,
				colormap name=myBlues,
				point meta min=0,
				point meta max=30,
				xtick=data,
				ytick=data,
				symbolic x coords={Imperative, Declarative, Hybrid},
				symbolic y coords={Qualify, Quantify, BDC, Explain, Localize},
				y dir=reverse, %
				xticklabel style={rotate=45, anchor=east},
				]
				\addplot[
				matrix plot*,
				mesh/cols=3, %
				nodes near coords,
				nodes near coords align={center},
				point meta=explicit,
				] 
				table [meta=z] {
					x y z
					Imperative Qualify 0
					Declarative Qualify 6
					Hybrid Qualify 1
					Imperative Quantify 26
					Declarative Quantify 5
					Hybrid Quantify 1
					Imperative BDC 10
					Declarative BDC 5
					Hybrid BDC 1
					Imperative Explain 19
					Declarative Explain 9
					Hybrid Explain 1
					Imperative Localize 15
					Declarative Localize 4
					Hybrid Localize 1
				};
			\end{axis}
		\end{tikzpicture}
		\caption{Conformance studies}
		\label{fig:technique-results-conformance}
	\end{subfigure}%
	\begin{subfigure}{.48\textwidth}
		\centering
		\begin{tikzpicture}
			\begin{axis}[
				width=0.8\textwidth,
				height=5cm,
				xlabel=Technique,
				ylabel=Results,
				colormap name=myReds,
				point meta min=0,
				point meta max=13,
				xtick=data,
				ytick=data,
				symbolic x coords={Imperative, Declarative, Hybrid},
				symbolic y coords={Qualify, Quantify, BDC, Explain, Localize},
				y dir=reverse, %
				xticklabel style={rotate=45, anchor=east},
				]
				\addplot[
				matrix plot*,
				mesh/cols=3, %
				nodes near coords,
				nodes near coords align={center},
				point meta=explicit,
				] 
				table [meta=z] {
					x y z
					Imperative Qualify 0
					Declarative Qualify 4
					Hybrid Qualify 0
					Imperative Quantify 2
					Declarative Quantify 9
					Hybrid Quantify 0
					Imperative BDC 0
					Declarative BDC 2
					Hybrid BDC 0
					Imperative Explain 1
					Declarative Explain 7
					Hybrid Explain 0
					Imperative Localize 1
					Declarative Localize 8
					Hybrid Localize 0
				};
			\end{axis}
		\end{tikzpicture}
		\caption{Compliance studies}
		\label{fig:technique-results-compliance}
	\end{subfigure}%
	\caption{Distribution of types of technique and types of results for the studies identified with Search A (\emph{conformance}) and Search B (\emph{compliance})}
	\label{fig:analysis_results}
\end{figure}

We see that most \emph{conformance} checking contributions (32 of 41) use \emph{quantitative measures}, i.e., numerical representations, to present the conformance checking results. One example for this is~\cite{rovaniDeclarativeProcessMining2015}, where a table detailing the counts of conformance violations and fulfilments is provided.
Few studies (seven in total) employ a \emph{qualification}, such as~\cite{chesaniAbducingComplianceIncomplete2016}, where a differentiation is made between strong compliance, conditional compliance, and non-compliance.
Further, less than half of the studies (i.e., 16) elaborate further on those measures by \emph{break-down and comparison}, for example~\cite{lemosUsingProcessMining2011} compares the most frequent variants and their conformity.
20 of 41 studies \emph{localize and show} violations or further results in, e.g., a process model. One example of this is~\cite{rinnerProcessMiningConformance2018}, where individual traces and occurring violations are considered and related to the general fitness value. 
However, more than two-thirds (i.e., 29 of 41) \emph{explain and diagnose} the observed violations in context and discuss potential causes and remedies, such as~\cite{stertzRoleTimeData2021}, where violations are explained by confirmed measurement errors and a previously undetected collision of robotic arms in a manufacturing process. 
Additionally, it should be noted that studies are not limited to one way of presenting the results, but usually use multiple representations throughout.

Moreover, we see that all but two of the conformance checking studies relying on imperative techniques use quantitative measures for representing results and no qualification. 
In contrast, half of the declarative applications employ qualitative measures and the other half quantitative ones.
Additionally, both imperative and declarative techniques (15 of 28, respective five of eleven) localize and show violations, but less often compared to qualifying or quantifying results. However, explanations and diagnoses appear to be relatively frequent (19 of 28, respective nine of eleven times) in declarative applications. As to hybrid applications, it is notable that all types of results are covered, and a focus seems to lie on qualifying and explaining diagnoses. Thus, it appears that the conformance checking technique chosen in the applications is related to the results and their contextualization and, therefore, should be chosen with care.

For the \emph{compliance} checking studies, we observe that around half of them (i.e., nine of 17) localize and show violations. We also see that almost two thirds (i.e., eleven) of the 17 studies quantify conformance. For instance, one of the two imperative compliance checking studies~\cite{weidlichEventBasedMonitoring2011} provides a numerical assessment of overall and average violations per type, and the share of violating process instances. However, in comparison to conformance checking techniques, we do not observe aggregate measures such as fitness values; instead, compliance violations are usually counted.
Four out of 17 studies provide a qualified conformance assessment (e.g.,~\cite{knupleschAFramework2017} where besides a quantification, for each state and each constraint, a violated/non-violated indicator is shown).
Less frequent are break-downs and comparisons, which are presented only in two of 17 studies (for example,~\cite{knupleschAFramework2017} where compliance information is shown for separate execution instances), as are explanations (eight out of 17 studies), such as~\cite{weidlichEventBasedMonitoring2011}, where an explanation for a particular type of constraint violation is provided. Notably, the only studies to break down and compare compliance are two studies that aim to demonstrate compliance checking techniques, perhaps in order to showcase the potential of compliance checking techniques for nuanced analyses.

\section{Discussion}
\label{sec:discussion}

Following the analysis of the studies that apply conformance checking and other data-driven techniques for regulatory compliance monitoring, we discuss and contextualize our findings along our research questions.

\subsection{RQ1 --- Identification of Existing Contributions} The SLR has identified 58 existing contributions that concretely monitor business processes for their regulatory compliance. We have found that the majority of \emph{conformance} checking studies are focused on the domain of healthcare, finance, and public services. For \emph{compliance} checking uses, we see various domains, such as healthcare and finance, with a slight skew towards trade and supply chain settings.
From a temporal perspective, the research interest seems to have shifted from investigating compliance checking towards conformance checking for regulatory compliance monitoring. Several reasons might exist for this, such as an increased attractiveness of process-mining related research due to specialized venues and outlets, or, as we will discuss further below, the need to demonstrate the applicability of conformance checking for regulatory compliance monitoring purposes.
Notably, we observe that a large share of conformance checking studies focuses on demonstrating and improving compliance monitoring techniques, especially in healthcare and finance. For the compliance checking studies we investigated, we also observe a larger focus on improving compliance monitoring techniques than on demonstrating their application or improving actual process compliance. While a wide range of domains is covered, to either improve or demonstrate techniques, or to improve process compliance, we see that regulations related to broader issues such as \emph{sustainability} (see e.g.~\cite{klessascheckUnlocking2024}) are not at all represented, neither from the conformance nor the compliance perspective. Given increasing societal demands for compliance with environmental objectives, we argue that future research on regulatory compliance of business processes could --- and should --- take a more active interest in this regard.

\subsection{RQ2 --- Characteristics of Existing Contributions} Next, we look at the steps necessary for regulatory compliance monitoring of business processes with conformance and compliance checking techniques. 

\paragraph{Characteristics of Conformance Checking Uses}

As to the characteristics of the identified conformance checking uses, we have underlined that the operationalization of regulations into prescriptive models is, so far, a manual process, that often is in need of expert knowledge for interpreting the regulations. There also seems to be a lack of reporting regarding the validation of prescriptive models, either due to a lack of access to expert knowledge, or due to the general difficulty in deriving them from relevant regulations in a sound way. This ties back to the point raised above, where, when using conformance checking techniques in particular for regulatory compliance monitoring, the prescriptive model must also be proven to be regulatory compliant~\cite{groefsemaUseConformanceCompliance2022}---which is not reported upon in more than two thirds of the applications. 
Contributions that aim at automatically creating prescriptive models and incorporate expert knowledge in a more explicit manner could provide a benefit here, and some research in this direction has already been started~\cite{saiDetectingDeviationsExternal2023,klessascheckUnlocking2024}.

In contrast to the literature review by Dunzer et al.~\cite{dunzerConformanceCheckingStateoftheart2019}, we see a wide range of process perspectives being assessed together. This illustrates the importance placed on data, time, and resource perspectives for regulatory compliance checking with conformance checking.
Further, a noticeable focus on imperative techniques exists in the investigated conformance checking studies. While domains such as healthcare appear especially suited for declarative techniques, there is significant potential to show how (and in how far) declarative conformance checking techniques are applicable for regulatory compliance monitoring in areas such as software development, IT service management or science. This might allow developing further guidance on when to apply which conformance checking techniques.
Almost all applications are tool-supported, meaning they employ established conformance checking tooling for the analysis of regulatory compliance, and only some are applied in an automated or real-time fashion. Moreover, we see that the choice of conformance checking techniques seems to be related to the overall results and visualization of insights. In general, we observe a focus on quantitative analyses and less often on explanations and qualifications in imperative applications, with declarative applications focussing additionally on qualitative analyses, as well as explanations and diagnoses. Localizations of violations and break-downs are less common. This seems to imply that declarative conformance checking applications revolve more around providing context and explanations to end users, whereas the main focus of imperative conformance checking lies in providing high-level numeric measures, and explanations for them. Consequently, the type of technique should be chosen depending on the regulatory constraints and the overall intended types of results. Finally, we see a focus on real-world process event data being used in the studies, although this data is rarely made available.

\paragraph{Characteristics of Compliance Checking Uses}

Characterizing identified compliance checking uses, we have shown that the operationalization of regulations is, similar to the conformance checking studies, done manually, while expert knowledge appears to be needed infrequently. This may be due to a focus on technique improvement in the studies we analysed. Similarly, prescriptive models are hardly validated. However, this may also be due to the fact that studies where technique improvement is the main objective do not necessarily require a validation, but rather just need to be plausible in showing their applicability. Perhaps it might also be comparatively easier to formalize declarative constraints in compliance checking approaches, since we also identify a noticeable focus on declarative techniques. 
We also see a wide range of process perspectives being assessed together, with almost all studies assessing at least three process dimensions together. This could mean that the types of techniques used in compliance monitoring approaches are very expressive regarding the process dimensions they can cover, especially declarative techniques. In terms of process event data, there is a prevalence of real-world data, whilst the data is rarely made available. Tool-based usage is observed almost exclusively, as well as a prevalence on compliance quantification and localization. Notably, break-downs and comparisons of compliance only happen in two studies, which aim to demonstrate usages of compliance checking.

\paragraph{Commonalities and Differences}

Comparing the two angles for monitoring regulatory process compliance, we note that compliance checking techniques are primarily declarative, while conformance checking techniques range from largely imperative, to declarative and a few hybrid ones. While conformance checking approaches have been described as ``unsuitable'' for monitoring process compliance due to their techniques' focus on an imperative perspective~\cite{hashmiAreWeDone2018}, we see a share of declarative conformance checking applications painting a more nuanced picture.

Another difference lies in the range of process perspectives: while both angles almost always take the control-flow dimension into account, compliance checking techniques generally consider more perspectives simultaneously, while for the conformance checking field we observe many uses where only one or two dimensions are analysed together. Thus, compliance checking approaches appear a degree more expressive in the process perspective they capture than conformance checking approaches, especially regarding temporal constraints. We conclude that, depending on whether the regulatory constraints under consideration pertain to multiple process dimensions or just one or two, conformance and compliance checking techniques differ in their applicability. In particular, if the constraints are more straightforward to express declaratively and relate to multiple process dimensions, declarative compliance checking techniques could be more useful.

Looking at the motivation behind the studies, we note another main difference: Compliance checking studies appear to be done more for technical reasons (i.e., for improving techniques), whereas demonstration and process improvement also play a role in conformance checking studies. We assume that the larger role of demonstration might be because conformance checking techniques are, as described above, seen in a less-than-ideal position for regulatory compliance monitoring and thus need to argue for their applicability in various settings.
The differing goals might also be linked to the high prevalence of healthcare and finance settings for conformance checking, since these two settings have useable and important regulations that may strengthen the case of conformance checking, whilst compliance checking studies might have a broader focus on domains, since they do not need to argue for their applicability.
Similarly linked to different motivations might be the fact that expert knowledge is less frequently mentioned as a requirement and prescriptive models are validated less often in compliance checking studies -- when improving techniques, it may be more important to have a prescriptive model and an application scenario that is sufficiently plausible, and not necessarily completely transferrable into a concrete real-world setting.

In terms of tooling, we observe that both fields make extensive use of tool-based applications and that there is a lack of automated/real-time applications; however, conformance checking studies largely use existing tools such as ProM, whereas prototypical implementations play a larger role in compliance checking studies. One useful line of future research, which we elaborate upon below, might therefore be to consolidate or unify existing compliance checking approaches.
A final difference can be observed in the results the two different angles provide: Since quantification lies at the core of conformance checking, we see numeric expressions of compliance quite frequently in conformance checking studies, as well as explanations and diagnoses (perhaps due to the high share of real-world studies that aim to demonstrate and improve processes). While we also see quantifications in compliance checking studies, these are usually simple counts of violations and not fitness scores common to conformance checking; additionally, localizations appear to be more common in compliance checking studies. From this, we conclude that the intended outcome of applying either compliance or conformance checking techniques for regulatory compliance monitoring should be considered when deciding on a specific approach.

However, common to both angles is a relatively low validation of prescriptive models, and the absence of a generalized approach for deriving these models from relevant regulations. From our analysis, we therefore suggest that future studies 1.) explicitly describe the \emph{source} of regulatory constraints and justify the choice for specific constraints; 2.) describe how these regulatory constraints relate to a specific business process under consideration; 3.) explicitly describe the design choices and reasoning behind their formalization into prescriptive process models or process constraints; 4.) provide an argument for how the formalization is a valid representation of the regulatory constraints; and 5.) be transparent regarding who was responsible for this formalization, and whether external expert knowledge was used during either formalization or validation. As a positive example,~\cite{WOS:000540648600033} provide a very comprehensive description of how process-specific constraints are derived from regulatory constraints, and how they are made to relate to a given event log.

While the prevalent use of real-world process event data we found in both conformance and compliance checking uses for regulatory compliance monitoring is certainly commendable, a concern is the lack of datasets that are actually made available. We understand that studies in which the actual regulatory compliance of business processes is improved could be unable to publicly share such data, due to e.g. NDAs. However, for purposes of transparency and reproducibility, we argue that future studies, especially those aiming to improve techniques for regulatory compliance monitoring, should make process event data available in some form.

\subsection{RQ3 --- Research Opportunities} Analysing existing contributions, we see the following \emph{research opportunities} (ROs) for future work, summarized in Table~\ref{tab:ros}:
First, the prevalent use of imperative and declarative techniques, both in conformance checking and compliance checking usages for regulatory compliance monitoring, hints at a potential for further investigation of hybrid techniques and their application for regulatory compliance monitoring (RO1). This is underlined by the fact that hybrid approaches might be able to produce a wider range of results and may use multiple process perspectives, which we posit could prove advantageous.
Second, in terms of automatization (RO2), we see a small number of contributions that go beyond using existing conformance and compliance checking tooling. This illustrates a potential for further research in the area of truly automated or real-time regulatory compliance monitoring with both conformance and compliance checking. 
Third, we also determine the potential for research of automated \emph{conformance} checking for regulatory compliance monitoring in other domains and processes beyond healthcare and manufacturing, which are not yet considered in great numbers by existing contributions (RO3). For example, monitoring compliance with sustainability regulations is so far a largely manual undertaking, that would benefit from demonstrations of data-driven techniques~\cite{klessascheckUnlocking2024}. 
Fourth, we identify a need for approaches and techniques that help in deriving regulatory compliant prescriptive models for both conformance checking and compliance checking, which are applicable across a wide range of domains; automating some of the necessary steps also would appear helpful in research and practice (RO4).
Fifth, we also see an opportunity for \emph{compliance} checking to be used for regulatory compliance monitoring in real-world cases, given the comparatively lower share of real-world studies of compliance checking. In contrast to conformance checking approaches, which have been amply demonstrated in various domains and real-world settings, we believe that \emph{demonstrating} compliance checking techniques applied in, e.g., the healthcare domain with validated prescriptive models, would allow further insights into the applicability of compliance checking techniques compared to conformance checking techniques (RO5).
Finally, given the high share of prototypical implementations in the regulatory compliance checking studies, we believe that an extensible framework for (potentially automated) compliance checking that provides generic interfaces for developing and implementing compliance checking techniques (similar to how ProM is used in the conformance checking studies we investigated) might be a valuable addition (RO6).

\begin{table}[ht!]
    \caption{Research opportunities of regulatory compliance monitoring}
    \label{tab:ros}
    \centering
     \renewcommand{\arraystretch}{1.5}
     \small
    \begin{tabularx}{\textwidth}{|l|X|}\hline
        Research Opportunity & Description\\\hline\hline
        RO1 & Uses of hybrid techniques for regulatory compliance monitoring.\\\hline
        RO2 & Automated / real-time compliance monitoring tooling.\\\hline%
        RO3 & Automated conformance checking in domains besides healthcare and manufacturing for regulatory compliance monitoring.\\\hline
        RO4 & Generalizable techniques for supporting/automating the creation of regulatory compliant prescriptive models for regulatory compliance monitoring.\\\hline
        RO5 & Further application of compliance checking for demonstration/process improvement purposes in additional domains.\\\hline
        RO6 & Extensible frameworks for (automated/real-time) regulatory compliance checking to reduce reliance on prototypical implementations.\\\hline
    \end{tabularx}
\end{table}

In addition to these impulses, we observe in relation to both conformance and compliance checking approaches for regulatory compliance monitoring that the role of expert knowledge is under-illustrated and not detailed upon. To derive a clear procedure for a more automatic operationalization of regulations, including the creation and validation of prescriptive models, more details are needed.
As we have shown, existing conformance checking studies focus on analysing and demonstrating the potential of applying conformance checking techniques for regulatory compliance monitoring. In contrast, compliance checking studies focus particularly on improving techniques. Therefore, it is unclear from the current state of the literature what the actual \textit{tasks} are that regulatory compliance monitoring is meant to solve when applied in practice. To develop useful regulatory compliance monitoring tooling (regardless of whether a conformance checking or compliance checking perspective is pursued), we therefore suggest a structured requirement analysis of these tasks. A first step into the direction of clarifying this has been taken in~\cite{rehseTaskTaxonomyConformance2025a}, where concrete tasks of conformance checking are elicited to inform more useful visual representations in conformance checking tools.

\subsection{Threats to Validity}
\label{sec:discussion-threats}
Notably, our work underlies some limitations which pose threats to its validity. Arguably, the search terms and criteria for our SLR, as well as the focus on only four databases, limit the relevant studies we were able to identify. However, by incorporating forward-backward searches and by adding studies through prior knowledge, we sought to limit the influence of this on our findings, as recommended for ensuring sufficient coverage~\cite{websterAnalyzingPrepareFuture2002,vombrockeStandingShouldersGiants2015}. Further, descriptive reviews such as the one presented in this study do not strive for (exhaustive) comprehensiveness, but instead conduct structured searches to identify a \emph{sufficient} body of works to characterize a broader field~\cite{pareSynthesizingInformationSystems2015}. With forward-backward searches and the use of expert knowledge, we have taken steps to ensure this sufficiency.

Further, we identified less than half as many relevant studies in Search B as in Search A, which limits the generalizability of our findings. Still, the majority of studies we filtered out for Search B either is not at all related to business processes, or focusses on proposing novel or improved techniques for regulatory compliance monitoring with no (empirical) evaluations or applications thereof. This may be due to a difference in what conformance checking-focussed studies and compliance checking-focussed studies perceive as a valuable contribution. Indeed, we assume that the field of process mining (and thus, conformance checking studies) is more data-driven than the compliance space we aimed to capture with Search B, leading to a higher prevalence of conformance checking over compliance checking studies. Additionally, the term compliance checking is used in many other fields that are not concerned with monitoring the regulatory compliance of business process executions.

Moreover, and perhaps the main issue to underline here, is an overall inconsistency regarding the compliance and conformance terminology: In the strictest sense, conformance checking only relates to the relation between a process model and an event log and only when specific conditions regarding the checked conformance relation are met, also prove regulatory compliance~\cite{groefsemaUseConformanceCompliance2022}. Similarly, compliance checking at runtime checks the conformance between event log and regulatory constraints to prove regulatory compliance~\cite{groefsemaUseConformanceCompliance2022}. Conformance checking approaches that include other perspectives beyond the control flow, as we have seen multiple times in our study, can be understood as \emph{simultaneously} checking compliance \emph{and} conformance~\cite{groefsemaUseConformanceCompliance2022}. Hence, the boundary between conformance and compliance checking seems blurry, since conformance checking is commonly understood as strictly assessing a relation between a process model and an event log, and  as applying process mining techniques so that from the relation between process model and event log information about regulatory compliance might be gained~\cite{groefsemaUseConformanceCompliance2022}. As a consequence, as argued by Groefsema et al., the terminologies have been used inconsistently in research, too~\cite{groefsemaUseConformanceCompliance2022}. In order to circumnavigate the difficulty in establishing whether a study did indeed check compliance or conformance, we instead assess them based on how they position themselves and how we found them in our two searches (either as compliance checking or conformance checking studies), and analyse them accordingly. This still allows us to compare how the two different viewpoints on data-driven process compliance monitoring differ in their usage.

Finally, the inductive coding we applied to the studies is subjective due to the judgment and careful reading required. We addressed this threat by following a thorough protocol when analysing the studies, taking detailed notes during analysis, and resolving uncertainties and disagreements through discussions. 

\section{Conclusion and Future Work}
\label{sec:concl}

To conclude, we identified relevant contributions that practically apply compliance monitoring techniques to the regulatory compliance of business processes. We investigated properties to characterize them and noted that existing compliance monitoring approaches are largely only tool-assisted., and that prescriptive models are always created manually and infrequently validated. Conformance checking approaches, in particular, are also reliant on expert knowledge. This underlines a potential for research in approaches that aid in the knowledge-intense operationalization of regulations, assess their compliance in an automated fashion, and provide detailed results beyond numeric assessments. We also highlighted commonalities and differences between compliance and conformance checking studies, showing that usages from those two angles differ to some degree in their goals, their techniques, and their process perspectives.
In the future, we plan to investigate novel ways of visualizing compliance violations in a way that is actionable and contextualizes violations with the corresponding regulations. Based on the SLR reported herein, we observe that supporting experts in the derivation of prescriptive models from regulations is a valuable field for future research, which we aim to address.

\backmatter

\section*{Declarations}

\paragraph{Competing Interests.} The authors have no competing interests relevant to this article to report.

\clearpage
\begin{appendices}

\section{Coding Table}\label{secA2}

\begin{table}[ht!]
    \centering
    \caption{Extracted categories and characteristics}
    \label{tab:categories_explained}
     \renewcommand{\arraystretch}{1.5}
     \tiny
    \begin{tabularx}{\textwidth}{|>{\raggedright\hsize=.16\hsize}X| >{\raggedright\hsize=.18\hsize}X >{\hsize=.66\hsize}X|}
    \hline
    Category & Characteristic & Description \\
    \hline\hline
    Domain & & Business domain of the process to which the study applies compliance monitoring\\\hline 
    \multirow{3}{8em}{Goals\\(non-\\exclusive)}
    & Improve Compliance & The study aims to improve a process to increase its compliance to certain regulations \\%\cline{2-3}
    & Improve Techniques & The study aims to improve compliance monitoring techniques, e.g., by extending existing ones or proposing new ones \\%\cline{2-3}
    & Demonstrate & The study aims to demonstrate compliance monitoring for a certain process or a certain domain \\\hline
    \multirow{2}{8em}{Prescriptive\\ Model Creation\\(exclusive)}
     & Manual & A prescriptive model is created through human interaction \\%\cline{2-3}
     & Automated & A prescriptive model is created automatically by software \\%\cline{2-3}
    \hline
    \multirow{2}{8em}{Data Origin\\(non-exclusive)}
     & Real-world & The event log data to which compliance monitoring is applied stems from real-world process executions \\%\cline{2-3}
     & Synthetic & The event log data to which compliance monitoring is applied is created synthetically \\%\cline{2-3}
    \hline
    \multirow{2}{8em}{Data Availability\\(non-exclusive)}
     & Publicly available & The event log data to which compliance monitoring is applied is publicly available\\%\cline{2-3}
     & Not available & The event log data to which compliance monitoring is applied is not made publicly available \\%\cline{2-3}
    \hline
    Prescriptive Model Validation & & Binary category, indicating whether the prescriptive model was explicitly validated\\
    \hline
    Expert Knowledge & & Binary category, indicating whether domain expert knowledge was explicitly required for applying compliance monitoring \\
    \hline
    \multirow{6}{8em}{Technique\\(exclusive)}%
     & Imperative & Compliance monitoring techniques that use an imperative process model as a prescriptive model are used\\%\cline{2-3}
     & Declarative & Compliance monitoring techniques which use a declarative prescriptive model\\%\cline{2-3}
     & Hybrid & Hybrid compliance monitoring techniques combine imperative and declarative techniques are used \\
    \hline
    \multirow{5}{8em}{Process Perspective\\(non-exclusive)}%
     & Control-flow & Compliance constraints related to the ordering of process activities \\%\cline{2-3}
     & Temporal & Compliance constraints related to temporal aspects of process activities\\%\cline{2-3}
     & Data & Compliance constraints related to data/documents used by the process activities\\%\cline{2-3}
     & Resource & Compliance constraints related to the resources executing the process activities \\
    \hline
    \multirow{6}{8em}{Application\\(non-exclusive)}
     & Manual & Compliance monitoring is performed entirely manually  \\%\cline{2-3}
     & Tool-supported & Compliance monitoring is performed by human interaction with tools (e.g. ProM)\\%\cline{2-3}
     & Automated & Compliance monitoring is fully automated by software and human interaction is not mandatory\\%\cline{2-3}
     & Real-time & Compliance monitoring is fully automated by software in real-time and human interaction is not mandatory \\
    \hline
    \multirow{4}{8em}{Results\\(non-\\exclusive)}\\
     & Qualify & Formal compliance evaluation without quantification (e.g., strong or weak compliance of a trace)\\%\cline{2-3}
     & Quantify & High-level indication with numbers about the compliance (e.g., fitness value, \# of violations etc.)  \\%\cline{2-3}
     & Break-down \& Compare & Additional charts/tables that further \emph{break-down} compliance indicators (e.g., average fitness or \# of violations) to \emph{compare} their distribution across multiple perspectives\\%\cline{2-3}
     & Localize \& Show &  Use of process diagrams/visualizations (e.g., Petri nets, BPMN or more general flowcharts) that \emph{localize \& show} violations or compliance evaluations\\%\cline{2-3}
     & Explain \& Diagnose & A textual description or other means of \emph{explaining \& diagnosing} the compliance evaluations or violations provided in the article\\
    \hline
\end{tabularx}
\normalsize
\end{table}

\end{appendices}

\clearpage
\pagebreak
\newpage

\bibliography{sn-article}%

\end{document}